\RequirePackage{fix-cm}
\documentclass[twocolumn]{svjour3}          % twocolumn
\smartqed  % flush right qed marks, e.g. at end of proof
\usepackage{xurl}
\usepackage{graphicx}
\usepackage{courier}
\usepackage[most]{tcolorbox}
\usepackage{amsmath,amssymb,amsfonts}
\usepackage{hyperref}
\usepackage{threeparttable}
\usepackage{marvosym} %For the \Letter symbol

\usepackage{longtable}
\usepackage{color, colortbl}
\definecolor{Gray}{gray}{0.9}
\usepackage{xcolor,colortbl}
\definecolor{LightCyan}{rgb}{0.88,0.9,0.9}
\usepackage{graphicx}
\usepackage{subcaption}
\usepackage{tikz}
\usepackage{soul}

\usepackage{amssymb}
\usepackage{xspace}

\newtcolorbox{shadedbox}{
	drop shadow southeast,
	breakable,
	enhanced jigsaw,
	colback=white,
	boxrule=0.80pt,
	left=0.3em,
	right=0.3em,
	top=0.1em,
	bottom=0.05em
}

\newcommand*{\ie}{i.e.,\@\xspace}
\newcommand*{\eg}{e.g.,\@\xspace}

\newcommand*{\Ls}{LLMs\@\xspace}
\newcommand*{\LLMs}{Large Language Models\@\xspace}
\newcommand*{\CG}{ChatGPT\@\xspace}

\newcommand*{\totPapers}{714\@\xspace}
\newcommand*{\noDupPapers}{707\@\xspace}
\newcommand*{\totAbs}{21\@\xspace}
\newcommand*{\totFinal}{14\@\xspace}

\makeatletter
\newcommand*{\etc}{%
	\@ifnextchar{.}%
	{etc}%
	{etc.\@\xspace}%
}
\makeatother
\newcommand*{\etal}{\emph{et~al.}\@\xspace}
\newcommand\revised[1]{\textcolor{black}{#1}}

\newcommand{\rqfirst}{\textbf{RQ$_1$}: \emph{How has existing research explored the application of LLMs in MDE tasks?}}

\newcommand{\rqsecond}{\textbf{RQ$_2$}: \emph{What strategies and methodologies have been developed to leverage LLMs in supporting MDE tasks? }}

\definecolor{verylightgray}{gray}{0.99}
\definecolor{lightgray}{gray}{0.92}
\definecolor{deepblue}{rgb}{0,0,0.5}
\definecolor{deepred}{rgb}{0.6,0,0}
\definecolor{deepgreen}{rgb}{0,0.5,0}
\definecolor{shadecolor}{gray}{0.9}

\definecolor{mygreen}{rgb}{0,0.6,0}
\definecolor{mygray}{rgb}{0.95,0.95,0.95}
\definecolor{myred}{rgb}{0.5,0,0}

\lstdefinestyle{JavaStyle} {
	backgroundcolor=\color{verylightgray},   
	commentstyle=\color{mygreen}, 
	breakatwhitespace=false,
	keywordstyle=\color{violet},
	language=Java,
	stringstyle=\color{blue},
	basicstyle=\scriptsize,
	showstringspaces=false
}

\lstdefinestyle{PythonStyle} {
	backgroundcolor=\color{white},   
	commentstyle=\color{deepgreen}, 
	breakatwhitespace=false,
	keywordstyle=\color{deepblue},
	language=Python,
	stringstyle=\color{deepgreen},
	basicstyle=\footnotesize\ttfamily,
	frame=single,
	showstringspaces=false }
	
\lstdefinestyle{searchstringstyle}{
	basicstyle=\ttfamily\scriptsize,
	captionpos=t,                    
	numbers=none,                    
	numbersep=5pt,                  
	showspaces=false,                
	showstringspaces=false,
	showtabs=false,                  
	tabsize=2,
	frame=single
}
\lstdefinestyle{pe}{
	belowcaptionskip=1\baselineskip,
	breaklines=true,
	frame=none,
	numbers=right,
	basicstyle=\scriptsize\ttfamily,
	keywordstyle = {\color{black}\bfseries},
	keywordstyle = {\bfseries},
	morekeywords={s,INST,SYS}, % Aggiungi qui altre parole chiave se necessario
	commentstyle=\itshape\color{purple!40!black},
	identifierstyle=\color{blue},
	backgroundcolor=\color{gray!10!white},
	emph={<s>, [INST]}, % Parole da mettere in grassetto
	emphstyle=\bfseries,
}

\definecolor{specialblue}{RGB}{66, 135, 245}

\begin{document}

\sloppy

\title{On the use of Large Language Models in Model-Driven Engineering}

%Grants or other notes

%\subtitle{Using generic transformations} % subtitle


\author{Juri~Di~Rocco \and 	  
	Davide~Di~Ruscio  \and
	Claudio~Di~Sipio \and		
	Phuong~T.~Nguyen  \and 	  
	Riccardo Rubei    
}

\authorrunning{Di Rocco, Di Ruscio, Di Sipio, Nguyen, Rubei} %\etal if too long for running head

\institute{Juri Di Rocco \at
	Universit\`a degli studi dell'Aquila, Italy \\ %67100 -- 
	%	Universit\`a degli studi dell'Aquila,	Via Vetoio 2, L'Aquila, Italy \\ %67100 -- 
	%	Tel.: +39-0862-433127\\
	\email{juri.dirocco@univaq.it}           %  \\
	\and	\Letter~Davide Di Ruscio \at
	Universit\`a degli studi dell'Aquila, Italy \\ %67100 -- 
	%		Universit\`a degli studi dell'Aquila,	Via Vetoio 2, L'Aquila, Italy \\ %67100 -- 
	%	Tel.: +39-0862-433735\\
	\email{davide.diruscio@univaq.it}
	\and
	Claudio Di Sipio \at
	Universit\`a degli studi dell'Aquila, Italy \\ %67100 -- 
	%	Universit\`a degli studi dell'Aquila,	Via Vetoio 2, L'Aquila, Italy \\ %67100 -- 
	%	Tel.: +39-0862-433735\\
	\email{claudio.disipio@univaq.it}		
	\and
	Phuong T. Nguyen \at
	Universit\`a degli studi dell'Aquila, Italy \\ %67100 -- 
	%	Universit\`a degli studi dell'Aquila,	Via Vetoio 2, L'Aquila, Italy \\ %67100 -- 
	%	Tel.: +39-0862-433127\\
	\email{phuong.nguyen@univaq.it}	
	\and
	Riccardo Rubei \at
	Universit\`a degli studi dell'Aquila, Italy \\ %67100 -- 
	%	Universit\`a degli studi dell'Aquila,	Via Vetoio 2, L'Aquila, Italy \\ %67100 -- 
	%	Tel.: +39-0862-433735\\
	\email{riccardo.rubei@univaq.it}	
}

\date{Received: date / Accepted: date}

\maketitle

\begin{abstract}
Model-Driven Engineering (MDE) has seen significant advancements with the integration of Machine Learning (ML) and Deep Learning (DL) techniques. Building upon the groundwork of previous investigations, our study provides a concise overview of current Language Large Models (LLMs) applications in MDE, emphasizing their role in automating tasks like model repository classification and developing advanced recommender systems. The paper also outlines the technical considerations for seamlessly integrating LLMs in MDE, offering a practical guide for researchers and practitioners. Looking forward, the paper proposes a focused research agenda for the future interplay of LLMs and MDE, identifying key challenges and opportunities. This concise roadmap envisions the deployment of LLM techniques to enhance the management, exploration, and evolution of modeling ecosystems. By offering a compact exploration of LLMs in MDE, this paper contributes to the ongoing evolution of MDE practices, providing a forward-looking perspective on the transformative role of Language Large Models in software engineering and model-driven practices.
\keywords{LLMs \and Generative AI \and Model-Driven Engineering}
\end{abstract}

\section{Introduction}\label{sec:intro}

%Automatically understanding and analyzing the modeling artifacts available in the modeling ecosystem of interest are preparatory for simplifying their management, including their exploration and evolution.

Model-Driven Engineering (MDE) promotes %is a software discipline that promotes 
the adoption of models to allow for the specification, analysis, and promotion of complex software systems~\cite{MDE}. A modeling ecosystem is made of available models, transformations, code generators, and a plethora of software tools.  With the integration of Machine Learning (ML) techniques, particularly in the context of modeling ecosystems, MDE has seen significant advancements lately~\cite{DiRuscio2023}. 

Language Large Models (\Ls), such as GPT-3.5, have demonstrated remarkable proficiency in understanding and generating human-like text~\cite{Martinez2022,10109345,Sauvola2024}. Very recently, there has been a dramatic increase in the number of applications of \Ls and pre-trained models (PTMs) in Software Engineering in general~\cite{NGUYEN2024112059}, and Model-Driven Engineering in particular. To name but a few, %For instance, 
LLMs have been widely used in testing~\cite{10440574}, code generation~\cite{10.1145/3643674}, qualitative research~\cite{DBLP:journals/ase/BanoHZT24}, summarization~\cite{10.1145/3593434.3593448}, or commit message generation~\cite{10433002}. However, to the best of our knowledge, there exists no work to provide a panorama view of current applications of \Ls in MDE, as well as to sketch a roadmap for future research directions in the domain.

Our work has been conducted to fill such a gap, providing an overview of the existing applications of \Ls in the MDE domain. \revised{We acknowledge that many techniques used in AI-supported SE could potentially be adapted for MDE tasks. However, this paper specifically aims to investigate the unique challenges and methodologies that arise when integrating LLMs within the MDE paradigm. The primary reason for excluding papers related solely to AI-supported SE is to maintain a clear and concentrated scope on MDE-specific applications. Essentially, including SE papers would broaden the scope too much, thus shifting the focus. Moreover, while textual encodings of models (\eg XML) provide a common ground for interoperability between SE and MDE, the context and objectives of their use in MDE are distinct. In MDE, these encodings are not merely textual representations but are used as inputs for model transformations, code generation, and other automated processes that are central to the MDE methodology. Therefore, we focused on studies that directly address these unique aspects of MDE.}
%
%In particular, the paper highlights the potential of \Ls in automating various tasks, including model specification, analysis, and development. 
The discussion involves the utilization of \Ls for automated classification of model repositories and the development of advanced recommender systems within modeling ecosystems. This paper builds upon the foundations laid by our previously published work entitled ``\emph{Machine Learning for Managing Modeling Ecosystems: Techniques, Applications, and A Research Vision}''~\cite{DiRuscio2023}. While the previous chapter explored the applications of traditional ML and Deep Learning (DL) in MDE, this paper goes one step forward, focusing on the utilization of \Ls to further enhance the capabilities of MDE.

To provide a comprehensive understanding, the paper delves into the technical intricacies of applying \Ls in MDE. In particular, it outlines the specific steps and considerations necessary to effectively support MDE tasks by means of  \Ls, ensuring seamless synergy between language understanding models and the complexities of model-driven systems.

\revised{In particular, we aim at answering the following research questions:
\begin{itemize}
	\item \rqfirst~By conducting a systematic literature review (SLR) \cite{kitchenham2004procedures}, we investigate \textit{(i)} the extent and manner in which current studies integrate LLMs into MDE tasks (LLM4MDE); and \textit{(ii)} the reverse scenario, where the MDE paradigm is employed to enhance the capabilities of LLMs (MDE4LLM);
	\item \rqsecond~We focus on the identification of the various approaches, frameworks, and tools proposed in the literature that utilize LLMs to facilitate different aspects of MDE. 
\end{itemize}}

In addition to the retrospective analysis, the paper sets forth a research agenda for the future of \Ls in MDE. It identifies key challenges and opportunities, proposing avenues for further exploration to maximize the potential of \Ls in enhancing the management, exploration, and evolution of modeling ecosystems. The envisioned roadmap encapsulates both the theoretical and practical aspects, paving the way for the deployment of LLM techniques in the MDE domain.

%By bridging the gap between language understanding models and the intricacies of model-driven systems, this paper contributes to the evolving landscape of MDE, presenting a forward-looking perspective on the transformative role of \Ls in shaping the future of software engineering and model-driven practices. Our ultimate aim is to provide researchers and practitioners with a practical guide to the applications of \Ls in Software Engineering and Model-Driven Engineering, allowing them to gain insights into the selection of suitable techniques and platforms.

%\dots

In this respect, the main contributions of our work are summarized as follows:

\begin{itemize}
	\item We provide a systematic literature review on the applications of \Ls in MDE.
	\item We discuss technical considerations for adopting \Ls to support the development of MDE tasks.
	\item Based on the current development in the domain, we present a research agenda organized with respect to the envisioned interplay of \Ls and MDE;
	
%	envision possible applications of \Ls and pre-trained models in various MDE tasks.
%	\item The data collected through our work has been published to facilitate future research.\footnote{\url{https://github.com/MDEGroup/LLMs-MDE}}
%	\item \PN{Something goes here.}
\end{itemize}
%\emph{``JQuery AJAX File Upload Error 500,''}

\textbf{Structure.} The paper is organized in the following sections. Section~\ref{sec:background} provides some background related to \Ls, prompt engineering, and hallucinations. Section~\ref{sec:integration} elaborates on  technical considerations for integrating \Ls into the MDE workflow. Afterwards, in Section~\ref{sec:SLR}, we present a systematic literature review on the applications of \Ls in MDE. Our research agenda is discussed in Section~\ref{sec:agenda}. The related work is then reviewed in Section~\ref{sec:relatedwork}. Finally, Section~\ref{sec:conclusion} sketches future work, and concludes the paper.

\section{Background in Large Language Models}
\label{sec:background}
%\PN{Add contents}
%\textbf{@Phuong}

As a base for further presentation, in this section, we recall basic concepts in the field of pre-trained and large language models, including prompt engineering, and hallucinations.

\subsection{The rise of \LLMs}

The recent months have witnessed a proliferation of pre-trained and large language models (\Ls). %Large Language Models (\Ls) represent a breakthrough in artificial intelligence, particularly in natural language processing (NLP). 
These models are characterized by their massive size, extensive pre-training on vast textual corpora, and sophisticated architectures based on deep learning techniques, notably transformer neural networks~\cite{NIPS20173f5ee243}. \revised{LLMs encompass various architectures, including both generative models like GPT (Generative Pre-trained Transformer) and masked language models like BERT (Bidirectional Encoder Representations from Transformers). GPT models are designed to generate text, making them suitable for tasks such as text generation and completion. On the other hand, BERT models are optimized for masked language modeling, where they predict missing words in a sentence, making them ideal for tasks such as text classification and token prediction.}

%\Ls, such as OpenAI's GPT (Generative Pre-trained Transformer) series, Google's BERT (Bidirectional Encoder Representations from Transformers), and Microsoft's Turing, are at the forefront of this paradigm shift in NLP.
%\todo{Comment 2.3 @Juri @Phuong}
\revised{\Ls are usually built on top of the Transformer architecture~\cite{NIPS20173f5ee243}, which consists of \emph{(i)} an Encoder to process the input text and generate a series of encoded representations; and \emph{(ii)} a decoder to use these representations to generate the output text. Moreover, there is the attention mechanism that allows LLMs to consider the entire context of a sentence when processing each word. In encoder-decoder models, cross attentions enable the decoder to focus on relevant parts of the input sentence when generating the output. 
%Thanks to this, %Through this exposure, 
\Ls learn to understand and generate text by capturing intricate patterns, semantic relationships, and syntactic structures inherent in human language.	%
During pre-training, the models are exposed to diverse text sources, ranging from books and articles to Web pages and social media posts.
Altogether, this equips LLMs with the ability to learn from a vast amount of text, and as a result, sophisticated artificial intelligence models, such as GPT-3 (Generative Pre-trained Transformer 3) or BERT (Bidirectional Encoder Representations from Transformers) %, have emerged as powerful tools, and 
they are capable of understanding and generating human-like text. %\Ls are pre-trained on vast amounts of textual data, allowing them to capture intrinsic language patterns and contextual nuances, semantic relationships, and syntactic structures inherent in human language. %The primary innovation of \Ls lies in their ability to learn rich, contextual representations of language through unsupervised pre-training on massive datasets. 
}

\revised{One of the key features of \Ls is their capability of generating informative answers, enabling them to produce coherent and contextually relevant text based on input prompts or cues~\cite{LIN2024122254}. This technical feature %generative aspect 
facilitates a wide range of applications, including language translation~\cite{ZIELINSKA2023101414}, text summarization~\cite{GOSWAMI2024111531}, question answering, sentiment analysis, and dialogue generation, to name but a few.} 
%In addition to their capability %prowess
%in natural language understanding, % and generation, 
%\Ls have also demonstrated %remarkable 
%versatility in domain-specific tasks. 
In %fields like 
 software engineering, \Ls offer unprecedented opportunities for code and text generation~\cite{10433002}, documentation automation, summarization~\cite{10.1145/3593434.3593448}, and bug detection~\cite{10109345,10176194}. By leveraging their deep understanding of programming languages and software development concepts, \Ls can assist developers in writing code more efficiently~\cite{BUCAIONI2024100526}, debugging applications, and comprehending complex codebases.
%In the realm of software engineering, these models find applications in a plethora of tasks, ranging from natural language understanding to code generation. They can assist developers in writing more efficient and accurate code, automating mundane programming tasks, and even aiding in the creation of documentation. 
The integration of \Ls into software development processes showcases their potential to enhance productivity and streamline various aspects of the software engineering lifecycle~\cite{10.1007/978-3-031-46002-9_23}. As these models continue to evolve, their impact on software engineering practices promises to be both profound and transformative. %
%In recent years, the field of software engineering has witnessed a transformative shift with the advent of large language models. %These sophisticated artificial intelligence models, such as GPT-3 (Generative Pre-trained Transformer 3) and BERT (Bidirectional Encoder Representations from Transformers), have emerged as powerful tools capable of understanding and generating human-like text. Large language models are pre-trained on vast amounts of textual data, enabling them to capture intricate language patterns and contextual nuances. 

%========================
%Despite their %remarkable 
%capabilities, \Ls also pose significant challenges, including \textit{computational requirements}, \textit{ethical considerations}, and concerns about \textit{biases} and \textit{fairness} in language generation. As the research community continues to push the boundaries of \Ls, addressing these challenges will be crucial to leveraging the full potential of these transformative technologies.
%========================

%Essentially, \Ls represent a groundbreaking advancement in natural language processing, offering unprecedented capabilities for understanding, generating, and manipulating text across a wide range of applications and domains. As \Ls continue to evolve, their impact on society, technology, and human-machine interaction is poised to be profound and far-reaching.

\subsection{Prompt Engineering}

%\DDR{It would be great to show simple and explanatory examples of prompt engineering, possibly in the MDE community. There is the paper ``On the assessment of generative AI in modeling tasks: an experience report with ChatGPT and UML'' that can be of inspiration.}

%~\cite{10.1145/3593434.3593448}, or commit message generation~\cite{10433002}

%\todo{Comment 2.4, Comment 3.4 @Phuong} 

\revised{To guide the behavior of \Ls during the deployment phase, %inference, 
prompt engineering is %involves 
the process of designing effective queries or input patterns. It is related to forming the input text, so as to yield the desired response or behavior from the model. The ultimate aim %goal 
of prompt engineering is to provide the model with context and constraints that steer its output towards the expected outcome. This involves specifying the task, providing relevant examples or instructions, and shaping the input to encourage the desired behavior, while minimizing undesirable outputs such as biases or inaccuracies in various domains such as %. Prompt engineering is crucial for leveraging the capabilities of \Ls in various applications, including 
text generation, question answering, or problem-solving~\cite{c0801d998c8c4722adf58b5d7eff447b,Taulli2023,10269066}.}
%
%We anticipate that 
There are the following possible applications of prompt engineering in software engineering~\cite{Sauvola2024}: %TODO: Look for any papers from David Lo's group on the related issues

\begin{itemize}
	
	\item \textit{Code Generation:} Developers make use of \Ls to generate code snippets based on prompts that describe the desired functionality or specifications. The use of prompt engineering is crucial while crafting input prompts that provide sufficient context and constraints to guide the model in generating accurate and syntactically correct code~\cite{roziere2023code}. %https://ai.meta.com/blog/code-llama-large-language-model-coding/
	
	\item \textit{Code Summarization:} \Ls can be employed to automatically summarize codebases, functions, or methods~\cite{10.1145/3551349.3559555}. In this context, prompt engineering aims to generate queries that allow the model to produce concise and informative summaries of code segments, while still preserving important details and functionality.
	
	\item \textit{Debugging and Problem-Solving:} Developers can employ \Ls in analyzing code snippets, identifying potential bugs, and suggesting solutions to common programming errors \cite{Fan_Gokkaya_Harman_Lyubarskiy_Sengupta_Yoo_Zhang_2023}. Prompt engineering in debugging tasks is to frame queries that describe the symptoms of the issue and provide relevant context to help the model diagnose and propose solutions.
	
	\item \textit{Documentation Generation:} LLMs can support developers in generating documentation for software projects, including function/method descriptions, API documentation, and usage examples \cite{Fan_Gokkaya_Harman_Lyubarskiy_Sengupta_Yoo_Zhang_2023}. In this case, prompt engineering is used to craft prompts that capture the key features and requirements of the software components to be documented.
	
%	\item \PN{Something goes here}. %Anything else?

\end{itemize}

Effective prompt engineering in software engineering requires an understanding of both the capabilities of \Ls and the specific requirements and challenges of software development tasks. It is necessary to iteratively refine prompts based on feedback, analyze model outputs, and incorporate domain-specific knowledge to ensure accurate and helpful responses from the \Ls. By means of prompt engineering techniques, developers can take advantage of the power of \Ls to streamline development workflows, improve productivity, and accelerate software development processes.

\begin{figure}[htbp]
    \centering
    \begin{subfigure}[b]{0.45\textwidth}
        \centering
        \includegraphics[width=\linewidth]{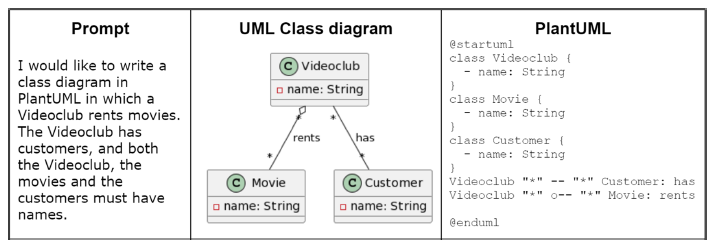}
        \caption{Domain modeling prompt (from C\'{a}mara \etal \cite{camara_assessment_2023})}
        \label{fig:domain_prompt}
    \end{subfigure}
    \hfill % This command puts space between the two subfigures if needed
    \begin{subfigure}[b]{0.45\textwidth}
        \centering
        \includegraphics[width=\linewidth]{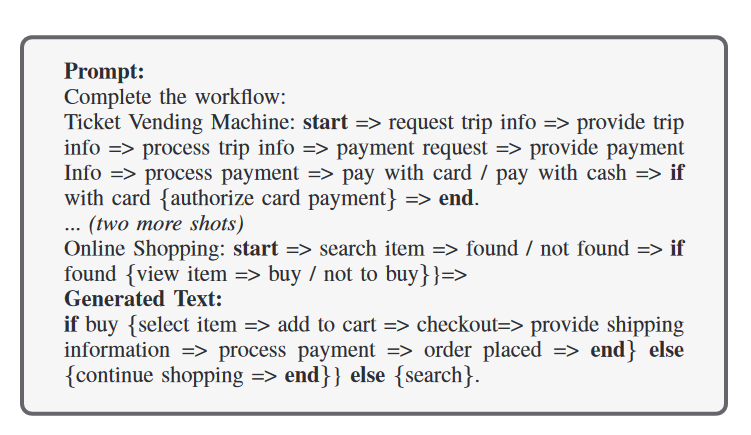}
        \caption{Model completion prompt (from Chaaben \etal \cite{chaaben2023towards})}
        \label{fig:completion_prompt}
    \end{subfigure}
    \caption{Examples of prompts defined for MDE tasks.}
    \label{fig:prompts}
\end{figure}

Considering the MDE domain, it is of a paramount importance to encode the information contained in the considered modeling artifacts. Figure \ref{fig:prompts} shows two explanatory examples of prompt engineering defined for assisting modelers in specifying two different kinds of model, \ie generic domain model (see Figure \ref{fig:domain_prompt}) and activity diagram (see Figure \ref{fig:completion_prompt}). It is worth mentioning that the first type of prompt, \ie the domain one, is a plain text that specifies the goal without considering the underpinning modeling elements. Meanwhile, the prompt shown in Figure \ref{fig:completion_prompt} embodies elements of the model since the goal is different, \ie model completion. In other words, prompt engineering strategy needs to be adapted to the corresponding MDE task. In the scope of this paper, we distinguish between the \textit{Raw text} prompting, \ie plain text prompts without model elements, and \textit{Template prompt engineering}, in which the natural language prompts are combined with model element. 

\revised{Several prompt engineering methods have been developed recently. For example, ReACT integrates logical reasoning with LLM outputs to improve decision-making processes \cite{yao2022react}. TreeOfThought employs a hierarchical approach to prompt design, enhancing the model's ability to handle complex tasks \cite{yao2023treethoughtsdeliberateproblem}. Self-consistency involves generating multiple responses and selecting the most consistent one, thereby improving the reliability of the outputs \cite{wang2023selfconsistencyimproveschainthought}}. 

\revised{In this section, we focus on chain-of-thought, few-shot prompting, and RAG because these methods are particularly relevant and widely applied in the context of our study on using LLMs in MDE tasks. Interested readers can refer to a recent paper \cite{amatriain2024promptdesignengineeringintroduction} presenting an introduction and advanced methods for prompt design and engineering.}

%\begin{figure*}[t!]
%	\centering
%	\begin{tabular}{cc}	
%		\subfigure[Domain modeling prompt by C\'{a}mara \etal \cite{camara_assessment_2023}]{\label{fig:domain_prompt}
%			\includegraphics[width=0.45\linewidth]{figs/prompt_domain_example_2.png}} &		
%		\subfigure[Model completion prompt by Chaaben \etal \cite{chaaben2023towards}]{\label{fig:completion_prompt}
%			\includegraphics[width=0.45\linewidth]{figs/completion_prompt_example.png}} 
%	\end{tabular} 
%	%	\vspace{-.2cm}
%	\caption{Examples of prompts defined for MDE tasks.} 	
%	\label{fig:prompts}
%	%	\vspace{-.4cm}
%\end{figure*}

\subsubsection{Chain of Thought}

\revised{Chain-of-Thought is a prompting technique designed to improve the reasoning capabilities of LLMs by breaking down complex problems into a series of intermediate steps~\cite{10.1007/978-3-031-47994-6_24,10.1007/978-3-031-40292-0_1,DBLP:conf/nips/Wei0SBIXCLZ22}. Unlike the traditional question-answering prompting technique, where each question is independent, Chain-of-Thought requires the model to understand and keep track of the context from previous interactions in the entire conversation. In particular, a series of questions are posed, each building upon the context established by the previous question and answer pair~\cite{LEE2024100213}. The ultimate aim is to evaluate the model's capacity for coherent, multi-turn dialogue and its ability to infer relationships and dependencies between questions and answers.}

\revised{Due to the need for long-term context retention and the ability to reason across multiple turns, Chain-of-Thought is particularly challenging for models. They serve as a benchmark for evaluating the performance of conversational AI systems and assessing their capabilities in handling complex, multi-turn dialogues. Developers use  Chain-of-Thought to identify strengths and weaknesses in conversational AI models and to guide further improvements in dialogue systems, natural language understanding, and reasoning abilities. By addressing the challenges posed by  Chain-of-Thought, systems can achieve more human-like conversational capabilities and improve their utility in various real-world applications, such as virtual assistants or customer service chatbots.}
 %, and dialogue assistants.

\subsubsection{Few-shot prompting}

%\todo{Comment 2.6 @Phuong} 

 \revised{In Natural Language Processing (NLP), few-shot prompting is a technique used to fine-tune or adapt large language models (LLMs) for specific tasks or domains using only a small amount of labeled data, the so-called ``few-shot'' dataset~\cite{CHEN2024111968}. In few-shot prompting, the model is provided with a prompt or example of the task along with a limited number of labeled examples, allowing it to generalize and learn the task quickly with minimal supervision~\cite{LI2024112002}.
The typical process, which is followed when doing few-shot prompting consists of the following activities~\cite{ZHOU2024127424}:}

\begin{itemize}
	\item \textit{Prompt Design:} A prompt is designed to provide the system with context and guidance about the task or domain that needs to be performed. The prompt serves as a template or instruction for the model to follow when generating responses or making predictions.
	
	\item \textit{Few-shot Dataset:} During fine-tuning, the system is fed with a small dataset containing labeled examples or instances of the task. The dataset may contain just a few examples, thereby yielding the term ``few-shot,'' but essentially, it should be representative enough to capture the key patterns and variations in the training data.
	
	\item \textit{Fine-Tuning:} The system is fine-tuned on the few-shot dataset by means of techniques such as gradient descent and back-propagation. While being trained, the system adjusts its parameters to minimize the discrepancy between its predictions and the ground truth labels contained in the few-shot dataset.
	
	\item \textit{Inference:} Once it has been fine-tuned, the system can be deployed for inference on new data or tasks related to the few-shot dataset. It uses the learned representations and parameters to make predictions or generate responses based on the input prompts or queries provided during inference.
\end{itemize}

\revised{It is important to note that fine-tuning, which involves training the model on a large amount of task-specific data to adjust its parameters, is not mandatory for few-shot prompting. Instead, few-shot prompting leverages the pre-trained capabilities of the model, using the examples in the prompt to guide its behavior and outputs.}

\subsection{\revised{Retrieval-augmented generation (RAG)}}

%Retrieval-augmented generation (RAG) %.Retrieval-augmented generation 

%\todo{Comment 2.7, Comment 3.9: @Juri @Phuong} 
RAG is a paradigm in NLP %natural language processing 
that combines elements of retrieval-based and generative models to improve the quality and relevance of generated text~\cite{lewis2020retrieval}. \revised{RAG focuses on enhancing the generation process by incorporating external knowledge through retrieval. In this way, it can be considered as a complement to prompt engineering, which deals with the crafting of the input prompts to guide the model's output.}
In retrieval-augmented generation, a generative model, such as an LLM, is augmented with a retrieval mechanism that fetches and incorporates relevant context or information from external sources during the generation process. Figure~\ref{fig:RAG} illustrates the RAG paradigm with %. The adoption of the RAG paradigm includes 
the execution of the following main phases:

%\color{blue}

\begin{figure*}[t!]
	\centering
	\includegraphics[width=0.8\linewidth]{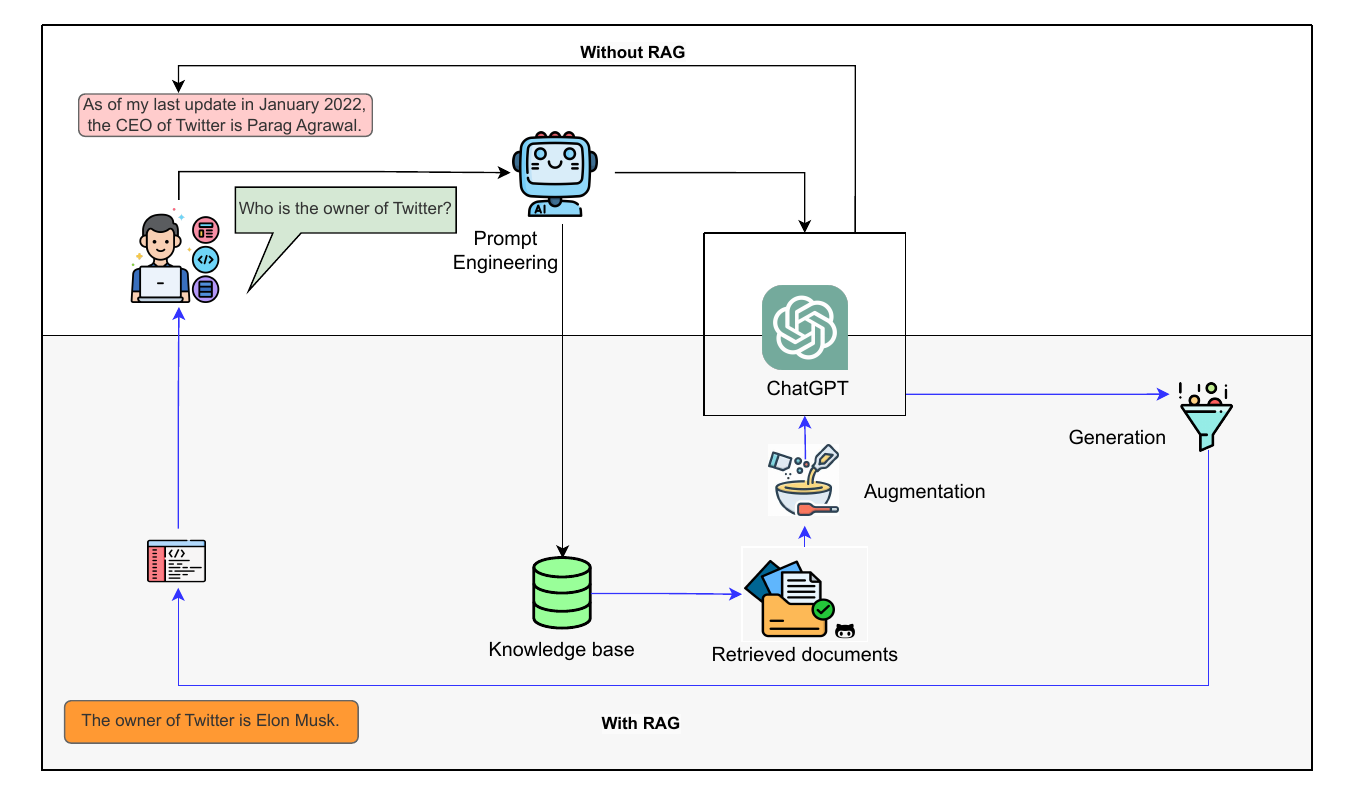}
	\caption{\revised{Retrieval-augmented generation (RAG).}}
	\label{fig:RAG}
\end{figure*}

\begin{itemize}
	\item \textit{Retrieval:} A retrieval mechanism is used to obtain and extract relevant context or information from a large corpus of text, knowledge bases, or external sources based on the input prompt or context. This retrieval can be done with various methods, \eg keyword matching, semantic similarity, or neural retrievers trained on large-scale text data.
	
	\item \textit{Augmentation:} The retrieved context is integrated with the input prompt to yield %and provided to the generative model as 
	additional input. % or conditioning information. %The generative model uses 
	This augmented input is used to generate text that is contextually relevant and coherent with respect to the retrieved information.

	\item \textit{Generation:} %The retrieved data is then augmented with context to enhance the prompt. 
	A generative model, such as a language model or neural network, is responsible for generating text based on the augmented %retrieved context and the input 
	prompt. The generative model leverages the retrieved information to enhance the relevance, coherence, and quality of the generated text. 
	This aims at producing a response augmented with the retrieved context. 
	
%	\item \textit{Generation Component:} A generative model, such as a language model or neural network, is responsible for generating text based on the retrieved context and the input prompt. The generative model leverages the retrieved information to enhance the relevance, coherence, and quality of the generated text.

\end{itemize}

\color{black}

%Figure~\ref{fig:RAG} illustrates the 

A typical application of RAG in practice is shown in Figure~\ref{fig:RAG}. Starting from the question: ``\emph{Who is the owner of Twitter?}'' then the answer generated by the corresponding LLM when there is no RAG (the upper part of the figure) is: ``\emph{As of my last update in January 2022, the CEO of Twitter is Parag Agrawal.}'' In fact, this is not the most updated knowledge as Twitter was sold to Elon Musk in October 2022.  
%The upper part of Figure~\ref{fig:RAG} represents the case when there is no RAG, and thus the answer is not yet updated. 
Meanwhile, in the lower part of Figure~\ref{fig:RAG}, we see that by consulting external sources, together with the initial query, the corresponding LLM is able to find a proper answer to the question, \ie ``\emph{The owner of Twitter is Elon Musk.}''

\revised{RAG enables an LLM to enhance its authenticity, diversity, and specificity in knowledge-intensive tasks by making use of the results mined from a specific database with either structured or unstructured multi-modal data~\cite{lewis2020retrieval,10.1145/3653081.3653102}. More importantly, RAG also helps to reduce factually inaccurate answers in text generation tasks~\cite{10.1162/tacl_a_00605}.} %
Thanks to this, RAG has found its applications in various NLP tasks, \eg text summarization, question answering, dialogue generation, and content creation~\cite{lewis2020retrieval}. By leveraging the complementary strengths of retrieval-based and generative models, RAG represents a promising method to enhancing the capabilities of AI-powered %language generation 
systems. In the scope of this paper, we are going to investigate the presence of RAG in MDE related applications.
 %\DDR{It would be great to a simple and explanatory example of RAG application (possible for each of the three phases above)}

%\PN{Something goes here}.

\subsection{Hallucination}

Hallucination in \Ls refers to instances where the model generates responses that are not grounded in the input context, or they are  not related to the conversation~\cite{10.1007/978-3-031-43458-7_34}. Hallucinations can occur when the model produces output being inconsistent with the input provided by the user, or when it generates information that is entirely fictional, nonsensical, or erroneous.
%
%Hallucination in Large Language Models (LLMs) refers to the phenomenon where the model generates responses that are not grounded in the input context or are unrelated to the conversation. Hallucinations occur when the model produces text that is inconsistent with the input provided by the user or when it generates information that is entirely fictional, nonsensical, or erroneous.
%
%Hallucinations happen due to 
There are the following reasons:

\begin{itemize}
	\item \textit{Contextual Understanding:} \Ls are trained on large amounts of text data and learn to generate responses based on patterns and associations present in the training data. However, sometimes they may fail to accurately understand or interpret the context of the conversation, leading to nonsensical responses.
	
	\item \textit{Lack of Grounding:} Hallucination can occur when the model generates responses that are not grounded in the input context or are disconnected from the topic or subject being discussed. This may result from limitations in the model's ability to retain and utilize context over long conversations or from errors in the generation process.
	
	\item \textit{Complexity of Language Understanding:} Natural languages are inherently complex, and \Ls may have difficulties in accurately capturing the nuances, ambiguity of human communications. In this respect, hallucinations may arise when the model misinterprets the meaning of the input or fails to recognize important contextual cues.
	
	\item \textit{Bias and Error Propagation:} Hallucinations can be triggered by biases contained in the training data. %, or errors propagated during the training. 
	If the training data contains biased or misleading information, the model may inadvertently produce hallucinatory answers reflecting these biases.
	
\end{itemize}

\begin{figure}[t!]
	\centering
	\includegraphics[width=0.55\linewidth]{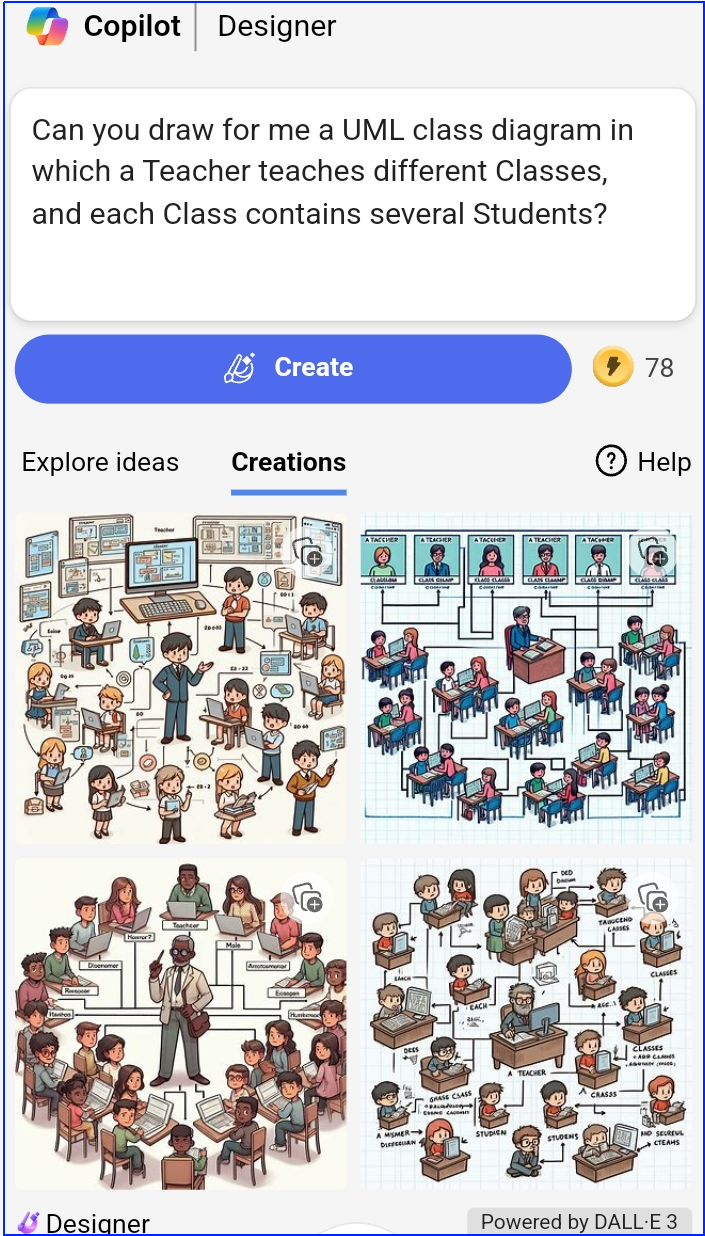}
	\caption{An example of hallucinations with DALL-E3.} % (from Wang \etal \cite{10.1007/978-3-031-53302-0_3})  (generated by Bing)
	\label{fig:HallucinationsDALLE}
\end{figure}

Figure~\ref{fig:HallucinationsDALLE} takes an example of hallucinations generated by Bing Copilot.\footnote{\url{https://www.bing.com/chat?q=Bing+AI&FORM=hpcodx}} With several attempts, we asked the engine to draw for us a UML diagram for Teachers, Classes, and Students. Surprisingly, Copilot misunderstood the query all the times and in the end, it just returned four pictures of physical classes with human teachers and students, and obviously this is not what we expected.

\begin{figure}[htbp]
	\centering
	\begin{subfigure}[b]{0.45\textwidth}
		\centering
		\includegraphics[width=\linewidth]{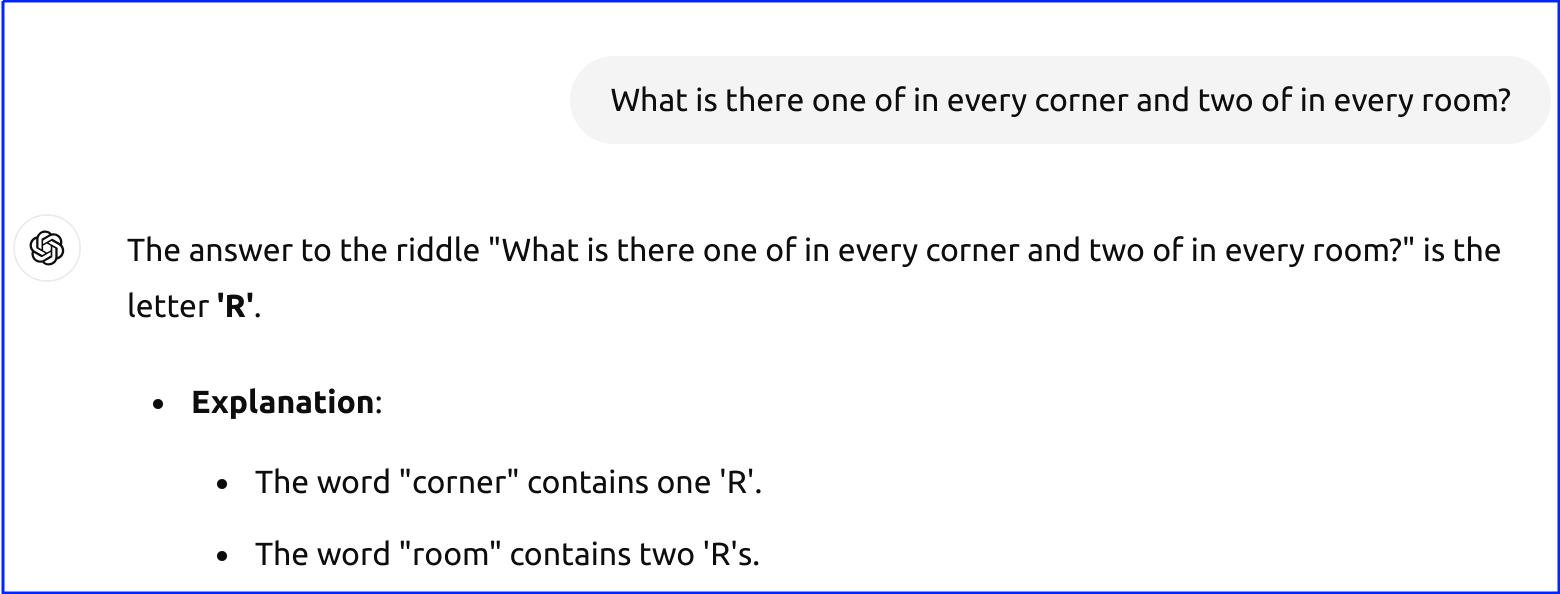}
		\caption{\revised{\CG gives the wrong answer by the first try.}}
		\label{fig:HallucinationsChatGPT_FirstTry}
	\end{subfigure}
	\hfill % This command puts space between the two subfigures if needed
	\begin{subfigure}[b]{0.45\textwidth}
		\centering
		\includegraphics[width=\linewidth]{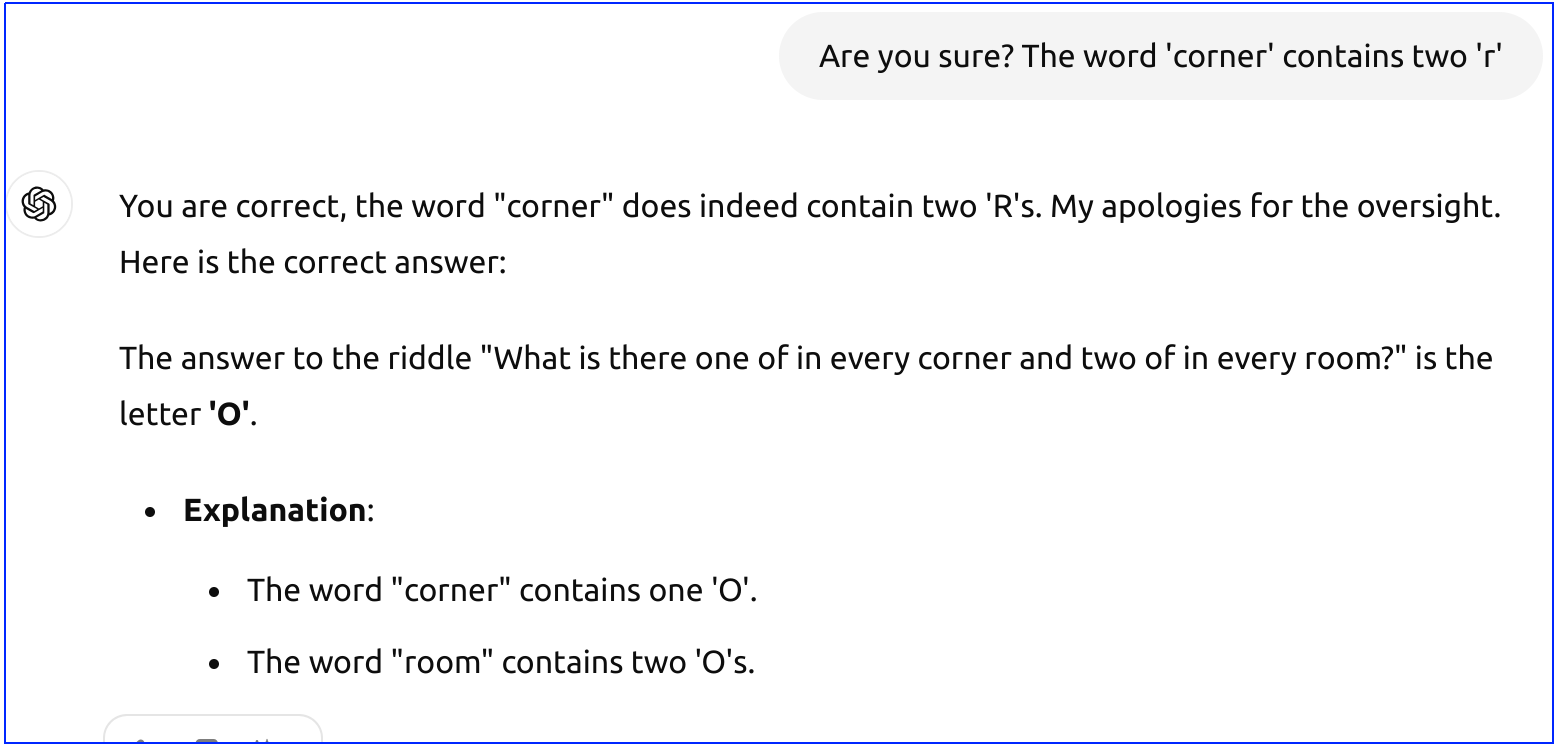}
		\caption{\revised{Once being corrected, it returns the right answer}}
		\label{fig:HallucinationsChatGPT_Corrected}
	\end{subfigure}
	\caption{\revised{An example of hallucinations with \CG.}}
	\label{fig:Hallucinations_ChatGPT}
\end{figure}

%\begin{figure}[h!]
%	\centering
%	\includegraphics[width=0.99\linewidth]{figs/Hallucinations_ChatGPT.png}
%	\caption{\revised{An example of hallucinations with \CG.}} % (from Wang \etal \cite{10.1007/978-3-031-53302-0_3})  (generated by Bing)
%	\label{fig:HallucinationsChatGPT}
%\end{figure}

\revised{In Figure~\ref{fig:Hallucinations_ChatGPT}, we show an example when \CG hallucinates with a text-to-text task. As in Figure~\ref{fig:HallucinationsChatGPT_FirstTry}, when being asked with the following question: ``\emph{What is there one in every corner and two in every room?}'' \CG gives an answer which reads ``\emph{The answer to the riddle ``What is there one of in every corner and two of in every room?'' is the letter 'R'}.'' Interestingly, this is not the right answer because the letter `R' appears twice in `corner' and only once in `room.' Only after being corrected, \CG has the right answer as shown in Figure~\ref{fig:HallucinationsChatGPT_Corrected}.}

%\begin{figure}[h!]
%	\centering
%	\includegraphics[width=0.99\linewidth]{figs/Hallucinations_ChatGPT_2.png}
%	\caption{\revised{\CG gives the right answer after being corrected.}} % (from Wang \etal \cite{10.1007/978-3-031-53302-0_3})  (generated by Bing)
%	\label{fig:HallucinationsChatGPTcorrected}
%\end{figure}

% \todo{Comment 2.8 @Davide}

So far, various strategies have been proposed to mitigate hallucination in \Ls~\cite{10.1007/978-3-031-53302-0_3}, including fine-tuning on domain-specific data, incorporating additional context or constraints into the generation process, and implementing post-generation filtering techniques to identify and remove hallucinatory responses.  %\DDR{Can we give some illustrative examples of Hallucination maybe by taking some from ~\cite{10.1007/978-3-031-53302-0_3}}
As shown later in this paper, we have not found any studies dealing with hallucinations in Software Engineering research, and this triggers the need for proper mechanisms to detect and deal with this phenomenon in Software Engineering and Model-Driven Engineering related tasks. %\todo{Comment 3.10 @Davide}

\section{Adopting LLMs to support MDE tasks: a technical overview}
\label{sec:integration}
%@Juri Show a running example to discuss all the different activities of the pipeline.
%In this section, we present an overview of the different activities that should be tackled to harness the capabilities of LLMs in supporting specific MDE tasks, i.e., enhancing model creation, manipulation, and understanding.

This section starts by sketching the main activities that are performed when employing \Ls to support MDE tasks, as well as discussing possible metrics for evaluating such types of applications. Afterwards, we provide a running example of generating UML diagrams via \Ls. \revised{The insights presented in this section are drawn from the authors' knowledge and experience in the topic.}

\subsection{Main activities to support MDE tasks with \Ls}\label{sec:process}

This section makes an overview of the main activities that need to be operated to leverage \Ls' capabilities to support specific MDE tasks, \eg  model generation, completion, and model management operations.
%
% In the following, we present a process to tailor \Ls to support MDE tasks. In particular, commonalities and variabilities specific to different projects or objectives have been conceptualized. For each study identified in Section~\ref{sec:background}, we analyzed common elements, differences, limitations, and challenges. By recognizing shared characteristics, we can more accurately identify trends or overarching patterns when utilizing \Ls to provide support in MDE tasks. Moreover, it is essential to analyze the disparities to comprehend \Ls' performance. 
%
%The analysis considers factors such as the type of MDE tasks, the specific pre-trained LLM model utilized, the method of prompting queries, the fine-tuning process with carefully selected datasets, the configuration of LLM hyper-parameters, and the chosen evaluation strategy.
%
Chen \etal~\cite{chen_automated_2023}  identified in their approach four steps to support the model completion task: \textit{(i)} problem formulation, \textit{(ii)} LLM architecture definition (including the model and hyper-parameter selection),  \textit{(iii)} model representation, and \textit{(iv)} post-processing of the results. We believe that such activities can be generalized and considered to be peculiar for any model management operation. In the following we discuss these four steps and consider an additional one related to the \textit{evaluation of \Ls}.

%\DDR{What about post-processing? I modified the following paragraphs according to the process activities proposed in ~\cite{chen_automated_2023} and listed above. In the previous version there was a completely disconnection between such a process and the discussed phases.}\JDR{I added a new paragraph for post-processing before ``evaluation of LLMs''}

%In the rest of this subsection we extend this process by considering more aspects, e.g., advanced techniques to enhance the performances of LLM and validation techniques.
 %Ultimately, we emphasize prevalent concerns, such as difficulties, obstacles, or constraints discovered in each study. These concerns encompass potential data gaps, threats to validity, or any other factors that may affect the reliability of the studies.
 
\paragraph{Problem formulation.} 

Identifying the specific MDE tasks to be supported by \Ls is a fundamental prerequisite before engineering the wanted support. \Ls, such as GPT, Llama, and their variants, offer a broad spectrum of capabilities from natural language processing to code generation and beyond. The vast potential of these models can only be effectively unleashed if the MDE tasks are clearly defined. For instance, tasks could range from generating UML diagrams from textual descriptions to automating the creation of domain-specific languages (DSLs)~\cite{camara_assessment_2023}. 
%
%Precisely identified tasks allow for more efficient use of LLMs by focusing their performances on generating relevant outputs. 
%	
%Task specificity helps reduce the computational overhead associated with more general applications and enhances the effectiveness of the solutions provided, ensuring they are directly applicable to the MDE tasks at hand. 
Moreover, clearly defined tasks enable the establishment of specific evaluation criteria, which are essential for assessing the performance of LLMs in supporting MDE tasks. For example, if the task involves classification of software models, the evaluation criteria might include standard metrics, \eg accuracy, completeness. 
%Precise task identification ensures that the undertaken evaluation methodology,  metrics are relevant and pertinent, facilitating an accurate assessment of the LLM's capabilities and areas for improvement. In addition, knowing the exact MDE tasks to be supported by LLMs aids in designing an effective integration workflow within the MDE ecosystem. 
Finally, having a well-defined conceptualization of MDE tasks impacts how other tools or processes will consume LLM outputs, the nature of the input data required by the LLM, and how feedback loops can be established to refine LLM performance continuously. 
%\DDR{Despite I removed a lot, the text does not go to the point about \textit{Problem formulation}. The question that raises after reading this paragraph is, \textit{ok, how to formulate the problem?}}
%\JDR{does the following paragraph solve your dubits?}
In other words, formulating a problem for \Ls requires meticulously mapping out the problem in terms of data requirements, expected LLM interactions, and desired outputs, to precisely match LLM strengths with MDE needs, as shown in the illustrative application presented in Section \ref{sec:llm-application}.

%In the follow we list the MDE tasks covered by the systematic literature review described in Section~\ref{sec:SLR}.

%\begin{itemize}
%	\item Model completion;
%	\item Model Search;
%	\item Model domain;
%	\item Model management operations.
%\end{itemize}

\paragraph{Modeling artifact representation.} 

Artifact encoding is a crucial preliminary step to employ \Ls into the MDE workflow under development. This process involves translating various MDE artifacts, such as models, metamodels, model transformations, and design documents, into comprehensible formats by \Ls. The aim is to bridge the gap between the highly structured, often graphical formats used in MDE and the text-based processing capabilities of \Ls.
Even though some artifacts, e.g., models and metamodels, are already encoded in textual formats, \eg XML, where the information is encapsulated in a way that is accessible to \Ls, for others, the adopted encoding strategies should face the loss of semantics, the complexity, and the ambiguity intrinsic of modeling artifacts.
%: Encoding must not significantly reduce the original artifact's semantic richness. Coping with this challenge requires careful consideration of what information is critical and how best to represent it textually or structurally.
%Complexity and Ambiguity: 

Due to their complexity, MDE artifacts can be challenging to encode, \eg in the case of model transformations that are specified with languages that include declarative and imperative programming constructs. Resolving ambiguities inherent in natural language or ensuring that the structured data accurately reflects the nuances of the original artifact adds another layer of difficulty. In addition, as artifacts evolve during the MDE process, it is crucial to ensure that the encoded representations are consistently updated to reflect changes. This consistency is vital for maintaining the integrity of the workflow and ensuring LLM-generated insights or artifacts remain relevant and accurate.

%Artifact encoding is a foundational step in leveraging the capabilities of LLMs for MDE tasks, requiring thoughtful consideration of the methods and formats used for encoding to ensure the effective integration of these powerful models into the MDE workflow. 

Encoding formats that are typically employed with \Ls include tree-based~\cite{weyssow2022recommending}, graph-based~\cite{shrestha_slgpt_2021}, EBNF models~\cite{chen_automated_2023}, JSON schemas~\cite{arulmohan_extracting_2023}, and textual forms such as plain\cite{camara_assessment_2023} and prompt-engineered text~\cite{10173990}. Each schema serves specific purposes, from preserving semantic integrity and modeling complex relationships to ensuring data structure and facilitating natural language processing. %The strategic application of these encodings enables \Ls to process, understand, and generate data across various domains, enhancing their utility in tasks that require nuanced comprehension and output generation.
%Moreover, encoded artifacts can serve as a lingua franca, enabling easier integration and interoperability between different tools and processes within the MDE ecosystem and beyond, including tools not traditionally designed to work with MDE artifacts.

%\smallskip
%\noindent
%\textit{Optimizing \Ls for MDE tasks} is essential to enhance the capabilities of \Ls in supporting MDE tasks. Techniques that can be employed to this end include \textit{RAG}, \textit{Knowledge Graph Construction}, \textit{Prompt Engineering}, and \textit{LLM Fine-Tuning}. This multifaceted optimization strategy ensures \Ls can be effectively integrated and tailored to meet the specific MDE task.

\paragraph{LLM architecture definition.} 

Recently, different language models have been released for targeting specific purposes, \eg Code Llama~\cite{roziere2023code} for coding, Llama and GPT  models~\cite{touvron2023llama,openai2023gpt4} for chatbot, to name a few. \revised{Only recently, research has been done to fine tune large language models for MDE tasks~\cite{shrestha_slgpt_2021,weyssow2022recommending}.} Thus, it is necessary to enhance the capabilities of general \Ls in supporting MDE tasks through different techniques that can be possibly employed, including  \textit{RAG strategies}, \textit{Knowledge Graph Construction}, \textit{Prompt Engineering}, and \textit{LLM Fine-Tuning} as discussed below.

\smallskip
\noindent
\textit{$\triangleright$ RAG strategies:} In the context of MDE, RAG can be instrumental in automating and enhancing various tasks. Customizing RAG to support MDE tasks involves several critical steps tailored to the specific requirements and challenges of MDE. First, it is necessary to rely on a comprehensive \textit{MDE knowledge base} that includes domain-specific models, code repositories, design patterns, documentation, and previous MDE artefacts. Furthermore, customizing the \textit{retrieval mechanism for MDE} could enhance the gathering of information to understand and prioritize information relevant to specific MDE task queries. %Train or fine-tune the retrieval component on MDE-specific queries and responses. This could involve creating a training dataset where pairs of MDE queries (e.g., "generate a UML diagram for a given set of requirements") and relevant artifacts (e.g., existing UML diagrams) are presented.
Actually, the current approaches reviewed in our systematic literature review do not make use of any RAG strategies, giving a possibility for further investigations.

\smallskip
\noindent
\textit{$\triangleright$ Knowledge Graph Construction:} It is pivotal for equipping \Ls with the context necessary for generating meaningful outputs. 
By structuring and connecting information derived from MDE artifacts, such as mega-modeling data sources \cite{di2020understanding,bezivin2004need}, knowledge graphs enhance LLMs' comprehension and analytical abilities. 
%None of the approaches reviewed in Section~\ref{sec:SLR} involve knowledge graphs.  %The construction of these graphs varies significantly in scope and complexity, depending on the requirements of the MDE tasks at hand. While some scenarios may only necessitate a basic graph linking elementary concepts, others might require an intricate network that encapsulates complex relationships and constraints, thereby enhancing the LLM's ability to provide context-aware solutions. 

\smallskip
\noindent
\textit{$\triangleright$ Prompt Engineering:} It emerges as a critical tool for effectively querying \Ls, guiding them towards generating the desired outputs. The art of crafting prompts involves a careful articulation of the tasks to the LLM, ensuring clarity and precision. Given the diverse nature of MDE tasks and the capabilities of different LLMs, prompt design is inherently variable. It demands iterative refinement to achieve optimal performance, with considerations for task complexity, the specific LLM in use, and the desired output format playing a pivotal role in this customization process. For this reason, the application of prompt engineering strategies have been deeply investigated in the approach proposed in the paper resulting from our systematic literature review. For instance, Chaaben \etal~\cite{chaaben2023towards} used few-shot prompt learning, which allows us to exploit these LLMs without having to train or fine-tune them on a specific domain or task, while Chen \etal~\cite{chen_automated_2023} conducted a comprehensive, comparative study of using \Ls for fully automated domain modeling, employing various prompt engineering techniques on a data set containing diverse domain modeling examples.

\smallskip
\noindent
\textit{$\triangleright$ LLM Fine-Tuning:} It further refines the model's alignment with MDE-specific requirements, significantly enhancing its accuracy and relevance to the tasks. Fine-tuning practices vary widely, from minimal adjustments based on a targeted dataset to extensive retraining on large, domain-specific corpora. This step is vital for ensuring that the LLM not only understands the intricacies of the MDE tasks but also produces outputs that are directly applicable and beneficial to them. Due to their resource-demanding nature, none of the reviewed approaches made us aware of these techniques for improving LLM performances.
	
%Including user feedback mechanisms  into the training loop to continuously improve the model's responses based on real-world usage and feedback~\cite{ouyang2022training,cui2023ultrafeedback}.

%It is worth noting that an user should considering these strategies, even in cooperation, when designing the usage of \Ls to support MDE tasks.

 \paragraph{Post-processing  of \Ls results.} A typical post-proces\-sing step transforms the output from LLMs into a specified model artifact format through a rule-based approach. When the output diverges from the expected format, the post-processor can adjust the output to preserve the validity of the generated answer, \eg domain models generated by the used LLM might need to be adapted when generated attributes lack a specified data type, and the post-processing can assign a default one to preserve the model validity~\cite{chen_automated_2023}.

 \paragraph{Evaluation of \Ls.} A tailored evaluation methodology is paramount for accurately assessing the model's performance and aligning it with the specific objectives when utilizing large language models (LLMs) for specialized MDE tasks. 
 %This tailored approach begins with clearly defining what constitutes success for the task and setting measurable goals that the LLM should achieve. For instance, if the task involves generating architectural models from requirements descriptions, the criteria might encompass aspects like model validity, correctness, and completion, ensuring that the evaluation reflects the unique demands of the application. 
  %
 %To effectively evaluate the practical effectiveness of Large Language Model (LLM) agents, it is crucial to utilize a diverse and representative test dataset that covers a broad range of scenarios and challenges likely to be encountered by the LLM. Evaluation methodologies vary to suit different testing needs and computational constraints, including common approaches like holdout validation for straightforward performance assessment, leave-one-out cross, and K-Fold Cross-Validation for a more comprehensive evaluation. However, computational demands may limit their applicability. Alternative methods like user studies and random subsampling provide valuable direct user feedback and performance metrics, emphasizing the importance of qualitative evaluations to identify usability issues and areas for improvement. Developers are encouraged to adopt an iterative evaluation approach, refining the model, dataset, and metrics based on testing feedback to enhance the LLM agent's performance and relevance.
 %
  Creating custom evaluation metrics for specific \Ls applications is essential, particularly when standard metrics such as accuracy and F1 score may not fully capture the model's effectiveness in specialized tasks. In tasks like model summary generation, utilizing metrics such as BLEU~\cite{callison2006re} and ROUGE~\cite{lin-2004-rouge} scores is crucial for assessing the quality of machine translation and summarization by comparing the LLM's outputs with manual summaries. Additional metrics, \eg  METEOR~\cite{meteor} and SIDE~\cite{mastropaolo2023evaluating}, can offer a deeper evaluation of these tasks, focusing on the nuances of language quality and summary relevance. This highlights the importance of tailoring evaluation metrics to reflect an \Ls performance accurately in its designated application.
   Furthermore, comparative analysis enriches the evaluation, placing the LLM's performance in context by benchmarking it against other baselines or methodologies addressing the same challenge. This highlights the LLM's unique strengths and weaknesses and uncovers potential areas for leveraging its capabilities more effectively.

 \subsection{Illustrative LLM application: from textual specifications to UML models}\label{sec:llm-application}

	%Because many approaches focus on model generation~\cite{chen_automated_2023,arulmohan_extracting_2023,camara_assessment_2023}, in the following we use the task of generating UML diagrams from textual requirements as an example application. %  can benefit from the usage of tailored LLM agents. 

%\todo{Comment 1.12 Juri}
In this section, we present an illustrative example showing an iterative process to devise an LLM to generate UML models from textual specifications. \revised{
	In the illustrative example, software engineers design complex systems to meet diverse requirements. Creating UML diagrams, such as use case and class diagrams, is often time-consuming and involves collaboration with multiple stakeholders. In recent years, various techniques have been explored to automate the generation of UML models~\cite{gulia2016efficient,deeptimahanti2009automated}. Inspired by the recent studies  of of Chen \etal~\cite{chen_automated_2023}, Arulmohan \etal~\cite{arulmohan_extracting_2023}, and C\'{a}mara \etal~\cite{camara_assessment_2023}, we show how an LLM can support the generation of UML diagrams based on a natural language description of the system. By following step-by-step guidelines, we explore the different technical choices an LLM engineer must consider. It is worth noting that we are presenting an explanatory example, which is incomplete on purpose to focus on its essentials.} %In our example we are will make use of typical python ml environment stack  including the most used python libraries. For instance, we will use pandas\footnote{\url{pandas.com}} fo data manipulation and analysis library, PyTorch\footnote{\url{URL}} can be used to train or fine-tuning models. Moreover, we will use hugging face libraries to handle LLaMA models on Hugging Face.
\revised{In our example, we will make use of a typical Python ML environment stack, including the most used Python libraries. For instance, we will use pandas\footnote{\url{https://pandas.pydata.org}} for data manipulation and analysis, and PyTorch\footnote{\url{http://pytorch.org}} can be used to train or fine-tune models. Moreover, we will use Hugging Face libraries\footnote{\url{https://huggingface.com}} to handle LLaMA models on Hugging Face Hub.}

LLaMA (via Hugging Face): Tools and models specifically for working with the LLaMA family of language models, integrated with the Hugging Face ecosystem..

%	
%	envision the potential engineering choices that a developer should consider when designing a tailored LLM agent to support the model generation. 
	%It is worth noting that it is 
%	Such a process is iterative, where each step could be refined until the desired generative quality is reached.
	%
	%The development of a tailored LLM agent for generating UML diagrams from textual requirements represents a sophisticated endeavor that leverages the cutting-edge capabilities of LLMs within the realm of MDE. 
	%

%, this process highlights the iterative activities of customizing \Ls to accurately and efficiently translate textual descriptions into structured UML diagrams.
	
First of all, developers must carefully analyze the specific UML models to be generated and understand the structure of the textual requirements, considering whether such requirements adhere to a particular template or format. This step is crucial for aligning the \Ls learning process with the task at hand, ensuring that the model can interpret and generate the desired models effectively.
	
UML models, inherently structured and XML-based, present a relatively straightforward scenario for encoding~\cite{lundell2006uml}. With its well-defined structure and widespread use in software engineering, XML offers a computer-readable framework that \Ls can navigate with relative ease. This inherent structure allows for the direct application of encoding strategies that leverage the XML-based nature of UML models, facilitating their comprehension and manipulation by \Ls without significant loss of semantic integrity or detail.
	
On the contrary, encoding text-based requirements demands a more sophisticated approach. Textual requirements, often expressed in natural language within a specified template~\cite{darif2023model}, involve a deeper analysis to select the most appropriate encoding method. This analysis is about %involves understanding 
the underlying structure and semantics of the template used and identifying the key components, such as entities, actions, and relationships, that are crucial for the accurate generation of UML models. The challenge lies in transforming this semi-structured natural language input into a format that retains its semantic richness while being accessible to \Ls for processing and analysis. 
To address this challenge, developers may consider various encoding strategies that span from semantic mapping to structured formats like JSON schemas or more advanced natural language processing techniques. Semantic mapping~\cite{chaaben2023towards}, for instance, involves creating representations that preserve the meaning of requirement specifications across different domains or contexts, ensuring that \Ls can accurately grasp the intent behind the textual descriptions. Alternatively, encoding requirements into JSON schemas provides a structured yet flexible format that can encapsulate the essential elements of the requirements, making them interpretable by \Ls.
	
	%Choosing the right LLM, such as Llama~\cite{touvron2023llama}, involves balancing computational resources, cost, update frequency, and availability. 
	%===
For this illustrative example, we decided to chose Llama,\footnote{\url{https://llama.meta.com/}} an open-source model suitable for deployment on personal infrastructures. Among the available models, Llama provides users with three main models, \ie \textit{Chat Model}, tailored for understanding and generating text in conversational contexts, 
\textit{Non-Chat Model}, a general-purpose model adept at a wide array of language-related tasks without specific optimization for conversational nuances; and the \textit{Code Model}, designed explicitly for programming-related tasks.
%In a research context focused on selecting an optimal language model for generating UML diagrams from textual descriptions, the developer chose Llama, an open-source model suitable for deployment on personal infrastructure. The developer decided on Llama, an model available for download and instantiation on personal infrastructure. 
%	
%Among the options were the Chat Model, tailored for understanding and generating text in conversational contexts; the Non-Chat Model, a general-purpose model adept at a wide array of language-related tasks without specific optimization for conversational nuances; and the Code Model, designed explicitly for programming-related tasks. 

Given the task's nature of this example, the \textit{Non-Chat model} was selected for its versatility across a broad spectrum of language-related tasks, avoiding the need for conversational context capabilities irrelevant to UML model generation. Finally, considering the balance between task complexity and available computational resources, a 13B parameter model size was identified as the ideal compromise. % This size efficiently meets the project's demands without exceeding the developer's computational capacity. 
%
%This example illustrates the critical importance of aligning model selection with project objectives and computational constraints, ensuring an effective and resource-efficient application of language models in specific tasks.
%
Listing~\ref{lst:mode-selection} shows an excerpt of code in Python to use the Llama pre-trained 13B non-chat model for the illustrative example.
%\revised{Line 1 of Listing~\ref{lst:mode-selection} import torch library to perform text generation using a trained model in an optimized inference mode (line 5 of Listing \ref{lst:hyperparameter}). LLama 2 pretrained model with 13B parameters have been retrieved from Hugging Face Model Hub in line 9 of Listing~\ref{lst:ModelSet}. AutoTokenizer imported in line 5 of Listing~\ref{lst:ModelSet} and initialized in line 2 of Listing~\ref{lst:hyperparameter} has been used to encode/decode data LLama model retrieved form from the Hugging Face Model Hub.}
\begin{lstlisting}[language=Python,numbers=left,caption=Llama model selection.,style=PythonStyle,label=lst:mode-selection, frame=single]
import torch
import huggingface_hub
import pandas as pd
from transformers import (
	AutoModelForCausalLM,
	AutoTokenizer,
	...	
)
model = AutoModelForCausalLM.from_pretrained(
"meta-llama/Llama-2-13b-hf",
...
)
\end{lstlisting}

\revised{Line 1 of Listing~\ref{lst:mode-selection} imports the \textit{torch} library to perform text generation using a trained model in optimized inference mode (as seen in Line 6 of Listing~\ref{lst:hyperparameter}). The LLama 2 pretrained model with 13B parameters was retrieved from the Hugging Face Model Hub in Line 9 of Listing~\ref{lst:mode-selection}. The \textit{AutoTokenizer}, imported in Line 5 of Listing~\ref{lst:mode-selection} and initialized in Line 2 of Listing~\ref{lst:hyperparameter}, was used to encode and decode data for the LLama model retrieved from the Hugging Face Model Hub.}

Upon the model selection, developers must investigate the configurations for various parameters pertinent to the chosen model. For instance, Listing~\ref{lst:hyperparameter} delineates the methodology for adjusting the \texttt{max\_new\_token} and \texttt{temperature} hyper-parameters for the response. The \texttt{max\_new\_token} parameter delineates the upper limit of new tokens that the model can generate in response to a specified prompt. Conversely, the \texttt{temperature} parameter modulates the randomness nature of the text generation process: lower values result in more deterministic outputs, whereas elevated values foster a heightened degree of diversity and innovation.
\revised{These two hyperparameters are commonly used across pretrained models and LLMs~\cite{zhu2024hot}, e.g., Hugging Face Transformers,\footnote{\url{https://huggingface.co/docs/transformers/main_classes/model?highlight=generate}} OpenAI GPT,\footnote{\url{https://platform.openai.com/docs/api-reference/chat/create\#chat-create-temperature}} and LLama.\footnote{\url{https://llama.meta.com/docs/llama-everywhere/}}
Listing \ref{lst:hyperparameter} provides an excerpt demonstrating how to set these hyperparameters on the LLama model.}	
\begin{lstlisting}[numbers=right, caption=Configuration of some Llama hyper-parameters.,style=PythonStyle, label=lst:hyperparameter, frame=single,language=python]
def summarize(model, text: str):
tokenizer = AutoTokenizer.from_pretrained(MODEL_NAME,use_auth_token=AUTH_TOKEN)
inputs = tokenizer(text, return_tensors="pt").to(DEVICE)
	...
with torch.inference_mode():
	outputs = model.generate(
		**inputs, 
		max_new_tokens=100, 
		temperature=0.0001)
...
\end{lstlisting}

Developers have the opportunity to build a knowledge base by leveraging both the requirements and model repositories to enhance the understanding and generation capabilities of \Ls. An interconnected graph structure, enriched with textual specifications alongside their corresponding UML models as highlighted in research by Huang et al.~\cite{huang2016enhancing}, acts as a contextual guide for \Ls. It enables the model to discern the relationships between textual descriptions and UML components effectively.

\begin{lstlisting}[numbers=right, caption=Searching UML models in ModelSet.,style=PythonStyle,label=lst:ModelSet,frame=single,language=python]
import modelset.dataset as ds
	
categories = ['health'] //More categories here
dataset = ds.load(MODELSET_HOME, modeltype = 'uml', selected_analysis = [])
modelset_df = dataset.to_normalized_df(min_ocurrences_per_category = 7, languages = ['english'])
df = df[df['category'].isin(categories)]
\end{lstlisting}

For this purpose, model repositories such as ModelSet \cite{modelset} can be utilized. ModelSet provides a comprehensive interface for querying UML models categorized based on specific criteria, as exemplified in Listing~\ref{lst:ModelSet}. This code snippet showcases the usage of the ModelsSet Python library~\cite{lopez2022using} to search for UML models related to the \texttt{health} category. Furthermore, sophisticated queries can be employed to retrieve models that encapsulate abstract representations of similar domains.

By carefully crafting prompts with detailed task descriptions, contextual hints, and domain-specific details, and by continuously refining these prompts based on performance feedback, developers can significantly enhance the generative capabilities of \Ls in converting textual requirements into UML models. %\todo{Comment 3.3 @Phuong}

For example, developers can enrich prompts with contextual clues extracted from the textual requirements. By incorporating these clues into the prompt, the LLM gains a better understanding of the specific aspects of the requirement that should be represented in the generated UML model. Furthermore, an iterative process of refining prompts can identify areas where the LLM may struggle with misunderstandings or require additional information or guidance on the structure and content of the desired UML model.

Creating prompt templates that can be adjusted for different types of UML model generation tasks provides a foundation for prompt engineering, ensuring consistency in how tasks are presented to the LLM while allowing for customization to specific requirements or domains. Moreover, establishing feedback mechanisms that involve expert reviews of generated UML models and user feedback in the prompt refinement process can help identify subtle nuances or complex requirements that may not be adequately addressed by the current prompts. Incorporating this feedback into prompt refinement improves the LLM's ability to produce high-quality UML models.

Listing~\ref{lst:prompt-engineering} show an excerpt of a possible prompt for the illustrative MDE task. The text enclosed by the special \texttt{<<SYS>>} tokens serves as context for the Llama model, guiding its responses based on our expectations. \revised{Line 2 contextualizes the conversation with the LLM with the following sentence: \textit{``You are a modeling assistant able to parse textual files containing requirements and generate UML sequence diagrams,''} specified between the \texttt{<<SYS>>} tokens. }This approach is effective because the same format was utilized during its training, incorporating a broad array of system prompts designed for diverse tasks. 

Throughout the course of the dialogue, each exchange between the human participant and the artificial intelligence entity is successively appended to the preceding prompt, demarcated by \texttt{[INST]} delimiters. To facilitate few-shot prompting, we employed the tokens \texttt{<s>} and \texttt{</s>} to demarcate sequences of one or more example questions and their corresponding answers. 
These examples are designed to guide the model's reasoning towards the expected direction. We utilized \{\{\dots\}\} as a placeholder intended to be substituted with specific contents, \eg \textit{UML.ecore} content is a placeholder for the XMI encoding of UML notation. Moreover, the user gives some models in one sequence to help generate valid UML models, \eg \textit{``Here a UML model that abstracts the class diagram for a voting system. \{\{UML class diagram instance\}\}"}.

\begin{lstlisting}[caption=An explanatory prompt for the illustrative MDE task.,label=lst:prompt-engineering,frame=single, style=pe]
<s>
[INST] <<SYS>>You are a modeling assistant able to parse textual files containing requirements and generate UML sequence diagrams <</SYS>> 
Here a set of textual sofware requirements {{software requirement}} 
[/INST] 
Can you provide me with the UML models in XMI format?
</s>
<s>
[INST] 
Here the UML models expressed in UML format: {{UML.ecore content}}
[/INST]
Can you provide me with a UML model instances in XMI format?
</s>
<s>
[INST] 
Here a UML model that abstracts the class diagram for a voting system: {{UML class diagram instance}}
[/INST]
Can you provide me with an instance of a UML model that includes a use case diagram?
</s>
<s>
[INST]
Here a UML model that includes at least one Use Case diagram for a voting system: {{UML use case instance}}
[/INST]
Which type of UML models would you like me to generate?
</s>
<s>
[INST] 
I would like to have a Use Case Diagram.
[/INST]

\end{lstlisting}

%``\emph{What is there one in every corner and two in every room?}''	

%\DDR{Listing 5 needs additional explanation. For instance, the text does not describe the meaning of the tag <s> </s> and of strings in \{\{\dots\}\} The sequence of the dialogue is not clear to me. Fist we see \textit{Here a UML model that includes at least one Use Case diagram for a voting system.}, and then we see the question \textit{Which type of UML models would you like me to generate?}} 
%\todo{Comment 1.14 Juri}

\revised{In particular, each discussion iteration delimited by the tags   \texttt{<s>}, \texttt{</s>} contributes to informing the LLM about the context of the discussion. For instance, while Line 3 provides the requirements, three iterations are simulated. First, the LLM is advise about UML specification to be used (Lines 5-10), then two instances of class and use case diagrams are provided within a defined domain, \ie voting system (Lines 11-22). Finally, the LLM asks which UML diagram the human would like to get from the given natular language requirements. This is just a demonstrative and not evaluated example. However, to the best of our experience, both the notations and model examples enhance the possibility of having valid models.}
	
	%The next critical step is fine-tuning the LLM with a corpus of template-based requirements and UML diagrams~\cite{elallaoui2018automatic}. This adjustment allows the LLM to utilize retrieved examples as context, aiming to produce UML diagrams that closely match new textual specifications. The process of fine-tuning is iterative, intending to optimize the model's performance for the specific task of UML diagram generation.
	
	%The LLM's effectiveness is evaluated using unseen textual specifications alongside their manually created UML diagrams, assessing the accuracy and relevance of the LLM-generated diagrams. This evaluation phase is key to identifying areas for further refinement, ensuring that the tailored LLM agent boosts productivity and enhances the quality of MDE artifacts.
	
Leveraging the availability of a comprehensive corpus of paired entities, i.e., template-based requirements alongside corresponding use case diagrams~\cite{elallaoui2018automatic}, the developers essayed on a fine-tuning process for the LLM. This fine-tuning is  designed to enable the LLM to contextualize and interpret the retrieved examples effectively. The ultimate goal was to empower the LLM to generate UML diagrams that accurately align with and reflect the essence of new textual specifications. This nuanced adaptation ensures that the LLM's outputs are not only syntactically aligned with the inputs but also semantically resonant with the underlying requirements.
	
For the evaluation phase, a rigorously curated set of previously unseen textual specifications, coupled with their manually crafted UML diagrams, was deployed as a benchmark to gauge the precision, accuracy, and overall relevance of the diagrams generated by the LLM. This critical assessment is a cornerstone for iterative refinement, enabling developers to systematically iterate on the LLM's performance. 
By analyzing the congruence between the LLM-gene\-rated diagrams and the manual benchmarks, developers can identify areas of improvement, \eg fine-tune the model's understanding and generative capabilities, and thereby enhance the efficacy and reliability of the LLM in automating the creation of UML diagrams from textual requirements. 

\section{Use of LLMs in MDE}
\label{sec:SLR}

%Literature review presenting the works done so far related to the application of LLMs in MDE. The dimensions that will be considered here are:
%
%- MDE tasks (coming from our book chapter + additional ones if needed)
%
%- Employed LLMs
%
%Moreover, I would like to have also a description of what are the general tasks that need to be performed use LLMs in MDE (as done for recommender systems)
%
%We will get different LLMs-related works that can be organized with respect to the supported tasks.
%
%Check the possibility to sell or the envision the idea that in the near future we will have small and task-specific LLMs. (Check if this can be moved/discussed to the research agenda.)
%

\begin{figure*}[t!]
	\includegraphics[width=.9\linewidth]{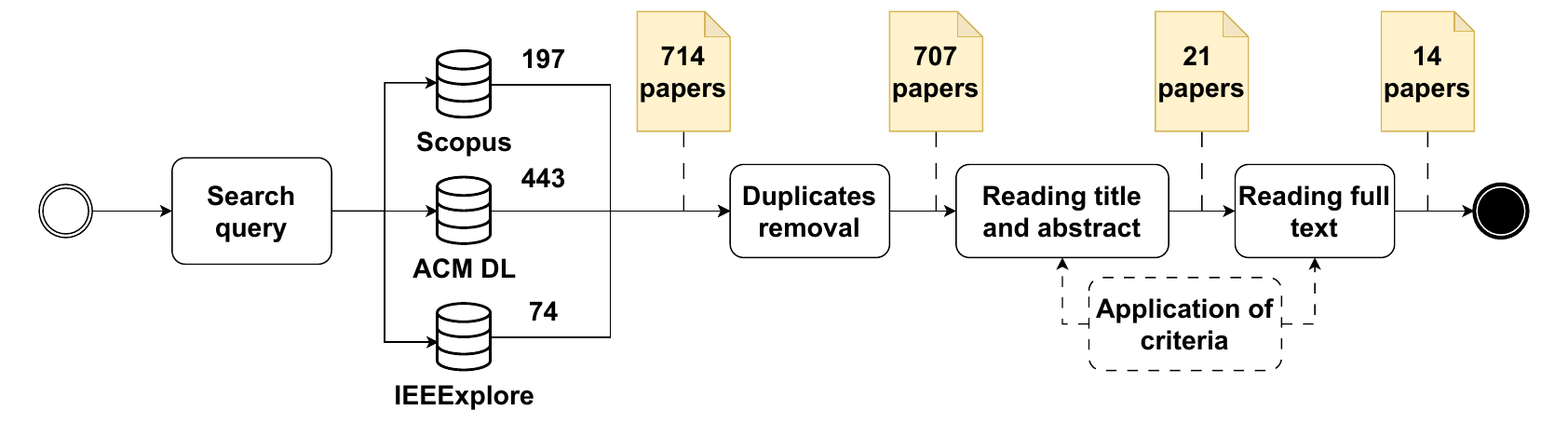}
	\caption{\revised{The process to retrieve relevant papers.}}
	\label{fig:slr_process}
\end{figure*}

\revised{This section aims to explore the current landscape of \Ls usage in model-driven engineering. To achieve this, we conduct a systematic literature review \cite{kitchenham2004procedures} across major scientific databases, seeking state-of-the-art studies. The process for retrieving relevant works is illustrated in Figure \ref{fig:slr_process}, incorporating three distinct digital libraries: Scopus,\footnote{\url{https://scopus.com}} ACM,\footnote{\url{https://dl.acm.org/}} and IEEE Xplore.\footnote{\url{http://ieeexplore.ieee.org/}} The query outlined in Listing \ref{lst:searchString} is employed during this systematic review.}

\begin{lstlisting}[label=lst:searchString,caption=\revised{The search string.},style=searchstringstyle]
( "system modeling" OR "software modeling" OR "model-driven engineering" OR "model-based software engineering" OR "model-driven development" OR "model-driven architecture" OR "model-driven software engineering" OR mdd OR mbse OR mde OR mda OR mdse AND ( large AND language AND model* ) OR llm OR llms OR pre-trained OR pre-trained language model*)
\end{lstlisting}

%``JQuery AJAX File Upload Error 500,''

Concerning the set of keywords, we searched for a specific set of tasks, \eg model transformation or model completion, plus synonyms of model-driven engineering, \ie MDD and MBSE. Since our work is focused on \Ls, we narrow down the scope of our research by using specific keywords used in this kind of model, \eg pre-trained or prompt engineering. Finally, we limited the search to recent papers, \ie those published from 2020. To ensure an unbiased selection process, we employed a rigorous approach. Two different authors independently evaluated all the papers, and the three senior co-authors thoroughly reviewed the entire selection process.

\revised{By executing the query on the three aforementioned digital libraries, we obtained a total of \totPapers papers. Subsequently, we filtered out duplicates that appeared in the selected sources, thereby reducing the number to \noDupPapers. From this refined list, we manually inspected titles and abstracts to identify papers aligning with our goals by applying inclusion and criteria described in Table \ref{tab:criteria}.}

       \begin{table}[htb!]
	\centering
	\footnotesize
	\caption{Inclusion and Exclusion Criteria.}
	\begin{tabular}{p{7.5cm}}
		\hline
		\multicolumn{1}{c}{\textbf{Inclusion criteria}} \\ \hline %\hline
		\revised{1. Papers that apply LLMs or pre-trained models to support MDE tasks (LLM4MDE)}     
		\\ \hline 
		\revised{2. Papers that apply MDE techniques, strategies or methodology to support LLMs definitions (MDE4LLM).}
		\\\hline
		\revised{3. Peer-reviewed papers published in high-ranking conferences or journals. They are identified based on established ranking systems such as the CORE Conference Ranking and the SCImago Journal Rank (SJR). Examples of such venues include, but are not limited to, ICSE, MODELS, and SoSyM.} 	
		\\ \hline
		\revised{4. Studies published over the last 5 years, \ie from January 2019 to July 2024}
		\\ \hline
		\multicolumn{1}{c}{\textbf{Exclusion criteria}} \\ \hline %\hline
		1. Foundation papers on LLMs or pre-trained models                        \\ \hline
		2. Papers not written in English.                        		\\ \hline
		\revised{3. Out-of-scope papers, \eg LLMs applied to generic SE tasks, MDE, MDE approaches applied to ML/DL networks}    \\ \hline		
	\end{tabular}
	\label{tab:criteria}
\end{table}

\revised{Following this inspection, we pinpointed \totAbs papers eligible for the next step, \ie reading the full paper. We eventually selected \totFinal studies\footnote{The interested reader can find the total number of papers and their details in the online appendix \url{https://github.com/MDEGroup/LLM4MDE-Appendix}} that leverage generative AI models to support modeling tasks. Subsequently, we meticulously analyzed each work to extract a set of relevant features characterizing them, \ie the employed \Ls, the supported MDE task, the managed artifacts, the encoding mechanisms applied to the input data, the prompt engineering strategy, the post-processing phase (if any) and the used evaluation strategies as shown in Table \ref{tab:comparison}.}

Weyssow \etal \cite{weyssow2022recommending} proposed a learning-based approach that leverages the RoBERTa pre-trained model to suggest relevant metamodel elements. The metamodels are encoded as structured trees to train the underlying model and obtain a textual sequential representation. Subsequently, a test set is generated using a sampling strategy relying on masking. %\DDR{What is it?}.
Essentially, masking is a technique employed in Natural Language Processing, where a portion of the input text is randomly modified~\cite{Wang2023}, allowing the model to learn to predict the masked text by relying on the contextual remaining words. The employed model is then used to predict missing elements and provide the modeler with insightful domain concepts. The results show that the employed model is capable of predicting the masked model items considering precision, recall and \revised{Mean Reciprocal Rank (MRR) metrics.}

Similarly, Chaaben \etal \cite{chaaben2023towards} support model completion using the GPT-3 model. Specifically, the model under construction is encoded using semantic mapping, \ie embedding the model elements as structured text in the prompt. Preliminary results computing traditional accuracy metrics on 30 models extracted from the ModelSet dataset~\cite{modelset} show that the few-shot approach can help modelers to complete static UML models, though there is still room for improvements, \eg the %overall 
accuracy can be increased by encoding non-natural language elements such as symbols and digits. \revised{In addition, completion of dynamic UML has not been evaluated rigorously.}

Shrestha and Csallner \cite{shrestha_slgpt_2021} proposed SLGPT, a fine-tuned version of the GPT-2 model to support the generation of graphical block-diagram models, \ie Simulink models. Initially, a curated training corpus of 400 valid open-source models is collected by combining a random model generator and a dedicated mining tool. The proposed approach then utilizes a breadth-first search (BFS) algorithm to preprocess the training models, \eg removing macros, default %configuration 
settings, and comments. SLGPT eventually generates a Simulink mo\-del by computing a probability mass function based on a well-known sampling technique and temperature. The conducted evaluation shows that SLGPT outperforms DeepFuzzSL, a baseline approach, in terms of generated structural properties. In addition, graph-based metrics computed on the generated sub-graphs demonstrate that SLGPT can generate adequate Simulink models using the internal validity checker component. 

%In \cite{chen_automated_2023}, the authors propose 
A framework to generate domain models in a textual format has been developed \cite{chen_automated_2023}, being compatible with various prompt engineering methods, and including a semantic scoring technique for evaluation. In addition, a dedicated post-processor module is devised to check the syntactic validity of the generated output using a rule-based method. The authors experimented with GPT-3.5 and GPT-4 using different prompt engineering methods, and conducted a detailed comparative evaluation with precision, recall, and F1-measure.
\revised{The results demonstrate that GPT-4 is powerful in understanding the domain according to the specified domain but not mature enough to completely automate model generation.}

%Similarly to the previous work, 
C\'{a}mara \etal \cite{camara_assessment_2023} explored the capability of GPT-3.5 in assisting modelers in their modeling tasks. Although GPT revealed to provide solid assistance with OCL expressions, it demonstrates many limitations in model generation. In particular, only for very specific domain like banking, the authors demonstrate a decent %level of 
precision. To evaluate the consistency of the generated model, a human-based evaluation has been carried out by creating 40 models belonging to 8 different domains. 

%In \cite{arulmohan_extracting_2023}, the authors investigated 

The ability of GPT-3.5 to extract information from requirements documents for model generation has been recently investigated~\cite{arulmohan_extracting_2023}. Specifically, the authors focused on the extraction from requirements concentrating on agile backlogs. In this work, 22 product backlogs and 1,679 user stories were used for extraction, and the evaluation consisted of comparing three approaches (Visual Narrator, GPT-3.5, and CRF) in terms of F1-score metric. Interestingly, the CRF implementation outperformed GPT-3.5.

Abukhalaf \etal \cite{10173990} assessed the Codex LLM's capabilities in generating OCL logical constraints defined on UML models. By manually creating prompts following a predefined template, the authors experiment different strategies, \ie basic prompts, zero-shots, and few-shots. The conducted evaluation on a dataset composed of 15 UML models and 168 specifications show that Codex obtains better results with few-shots technique in terms of validity score and accuracy even though the program repair techniques can increase the naturalness of the generated constraints. \revised{The same authors proposed PathOCL \cite{10.1145/3650105.3652290}, an approach based on GPT-4 model to support the generation of OCL rule using chuncking technique to overcome the token limitation issue. Afterward, the generated prompt are ranked according to well-know similarity functions, \ie Cosine and Jaccard. 
 In particular, the comparison with Codex model shows that GPT-4 is more efficient in generating OCL constraints using the augmented prompts in terms of the considered metrics, \ie Correctness and Validity.}

 \revised{Ahmand \etal \cite{ahmad_towards_2023} applied ChatGPT to support the generation of software architecture from textual requirements. First, the initial requirements are enforced by a continuous dialogue between chatGPT and the human architect. Afterward, a PlantUML diagram is generated by exploiting three well-founded architecting activities. ChatGPT is eventually used to evaluate the generated software architecture using the SAAM methodology \cite{10.1109/TSE.2002.1019479}.}

\revised{%In \cite{saglam_automated_2024}, the authors exploit 
ChatGPT has been used to assess plagiarism in modeling assignments~\cite{saglam_automated_2024}. First, a generic AI-based plagiarism detector has been developed by relying on an NLP-based plagiarism detector composed of four different phases, \ie tokenization, normalization, pairwise matching, and similarity calculation. Afterward, ChatGPT is instructed to replicate human assignment by using two different prompt techniques, \ie zero-shot asking for a full-generation and few-shots using obfuscation on existing models. By relying on an existing dataset of EMF-based metamodels, the conducted evaluation reveals that ChatGPT is not able to produce correct assignments even though using obfuscation improves the generation capabilities in terms of plagiarism metrics.}

\revised{Kulkarni \etal \cite{kulkarni_toward_2023} proposed a methodology to integrate GPT-4 and the development of digital twin models. First, the human designer specifies the system requirements using user stories directly in the prompt. Afterward, the prompt is iteratively refined using the Goal-Measure-Lever (GML) metamodel, in which the modeler can define the goals and sub-goals of the final system. The approach can eventually generate an enhanced model that can be transformed into DSL specification using MDE standard technique.}

\revised{Bertram \etal \cite{10.1145/3567512.3567534} exploited GPT-3 models to translate textual requirements in DSL specification in the context of Advanced Driver Assistance Systems (ADAS). Starting from unstructured textual requirements, the authors employ the few-shots prompt technique to derive formal rules used in the DSL specification. To evaluate the approach, the authors conducted an early validation using an Adaptive Light System as the motivating scenario.}

\revised{Apvrille and Sultan \cite{modelsward24} employed ChatGPT to complete structural and behavioral SysMl models. Built on top of the online TTool framework, the proposed approach first processes the user query composed of the partial SysMl and the domain knowledge encoded as a JSON request. Afterward, ChatGPT exploits the augmented query to enhance the model's generation. The TTool framework eventually extracts the GPT's response and delivers it to the modelers using a feedback mechanism to post-process any syntax error. The evaluation shows that the proposed framework slightly outperforms students in the modeling tasks, even though the results are worse when complex specifications are considered.}

\revised{Chen \etal \cite{10260905} exploited GPT-4 to guide the creation of goal models in the context of requirement engineering. Given the input specified using textual goal-oriented requirement language (TGRL), the authors exploit zero-shot and few-shot prompting to instruct GPT in completing goal models using two different types of questions, i.e., open and closed. In addition, each prompt contains a syntax description to detail the application context with dedicated TGRL tags. Finally, interactive feedback has been used to improve the results. The experiment conducted on two different use cases, \ie Kids Help Phone and Social Housing, demonstrates that GPT-4 is effective in specifying the goal models even though it fails to handle complex requirements. In addition, feedback can improve the overall results, although they may introduce model errors in certain cases.}

 \begin{table*}[t!]
    \centering
    \scriptsize
    
    \caption{\revised{Comparison of existing LLM4MDE approaches}.}
    \label{tab:comparison}
    \resizebox{\textwidth}{!}{
        \begin{threeparttable}
        \begin{tabular}{|p{2.5cm}|p{1.8cm}|p{1.8cm}|p{1.8cm}|p{1.8cm}|p{2.0cm}|p{1.8cm}|p{1.8cm}|p{1.8cm}|} % Adjusted the column widths
            \hline
            \textbf{Approach} &
            \textbf{Underpinning model} &
            \textbf{Modeling task} &
            \textbf{Modeling artifact} &
            \textbf{Artifact encoding} & % Original position
            \textbf{PE\tnote{1} Encoding} & % New column
            \textbf{PE strategy} & % Swapped position
            \textbf{Post-processing} & % New position
            \textbf{Evaluation} \\ \hline            
            Chaaben \etal \cite{chaaben2023towards} & GPT-3 & Model completion & Static and dynamic models & None & Template PE & Few-shots & None & Precision, Recall \\ \hline
            Weyssow \etal \cite{weyssow2022recommending} & RoBERTa & Model completion & Metamodels & Tree-based & N.A.\tnote{5} & N.A. & None & Precision, Recall, MRR\tnote{4} \\ \hline
            Shrestha and Csallner \cite{shrestha_slgpt_2021,shrestha_harnessing_2023} & GPT-2 & Model search & Simulink models & Graph-based & Raw Text & N.A. & Validity Checker & Graph-based metrics \\ \hline
            Chen \etal \cite{chen_automated_2023} & GPT-3.5, GPT-4 & Model Generation & Domain models & EBNF models & Raw Text & Zero and few-shots, CoT\tnote{2} & Rule-based & Precision, Recall, F1-score \\ \hline
            Arulmohan \etal \cite{arulmohan_extracting_2023} & GPT-3.5 & Model Generation & UML models & JSON schema & Raw Text & Rapid prototyping & JSON parser & F1-score \\ \hline
            Camara \etal \cite{camara_assessment_2023} & GPT-3.5 & Model Management Op. & UML models & None & Raw Text & None & None & Scenario-based \\ \hline
            Abukhalaf \etal \cite{10173990} & Codex & Model Management Op. & UML models & PlantUML & Template PE & Zero and Few-shots & Manual & Accuracy and validity \\ \hline       
            \revised{Abukhalaf \etal \cite{10.1145/3650105.3652290}} & \revised{GPT-4, Codex} & \revised{Model Management Op.} & \revised{UML models} & \revised{PlantUML} & \revised{Template PE} & \revised{Few-shots} & \revised{Manual} & \revised{Similarity\tnote{3}, Correctness, Validity} \\ \hline

			\revised{Ahmand  \etal \cite{ahmad_towards_2023}} & \revised{ChatGPT\tnote{6}} & \revised{Model architecture} & \revised{UML models} & \revised{PlantUML} & \revised{Raw text} & \revised{CoT} & \revised{None} & \revised{SAAM methodology \cite{10.1109/TSE.2002.1019479}} \\ \hline

			\revised{Sağlam \etal \cite{saglam_automated_2024}} & \revised{ChatGPT\tnote{6}} & \revised{Model search} & \revised{Metamodels} & \revised{EMF-based models} & \revised{Obfuscated models} & \revised{Zero and Few-shots} & \revised{None} & \revised{Plagiarism metrics} \\ \hline

            \revised{Kulkarni \etal \cite{kulkarni_toward_2023}} & \revised{GPT-4} & \revised{Model completion} & \revised{Metamodels} & \revised{GML model} & \revised{Raw Text} & \revised{CoT} & \revised{None} & \revised{Scenario-based} \\ \hline    

            \revised{Bertram \etal \cite{10.1145/3567512.3567534}} & \revised{GPT-3} & \revised{DSL requirement} & \revised{ADAS requirements} & \revised{DSL rules} & \revised{RawText} & \revised{Few-shots} & \revised{None} & \revised{Scenario-based} \\ \hline

            \revised{Apvrille and Sultan \cite{modelsward24}} & \revised{GPT-3.5} & \revised{Model completion} & \revised{SysMl models} & \revised{JSON schema} & \revised{JSON request} & \revised{RAG} & \revised{Feedback mechanism} & \revised{Time computation} \\ \hline    

            \revised{Chen \etal \cite{10260905}} & \revised{GPT-4} & \revised{Goal modeling} & \revised{TGRL models} & \revised{Structured TGRL} & \revised{Template PE} & \revised{Zero and Few-shots} & \revised{Feedback mechanism} & \revised{Scenario-based} \\ \hline

    \end{tabular}
    
    \begin{tablenotes}
        \item[1] PE  = Prompt engineering
        \item[2] CoT = Chain of Thoughts
        \item[3] The similarity has been assessed with Cosine and Jaccard distance
        \item[4] MRR = Mean Reciprocal Rank
        \item[5] N.A. = Not applicable
        \item[6] GPT version not specified
    \end{tablenotes}
    
\end{threeparttable}}
\end{table*}

%From the examined approaches, we extract a feature model shown in Figure \ref{fig:features} that represents the high-level concepts of the two application domains, \ie LLMs and MDE. 

\begin{figure*}
	\centering
	\includegraphics[width=.7\linewidth]{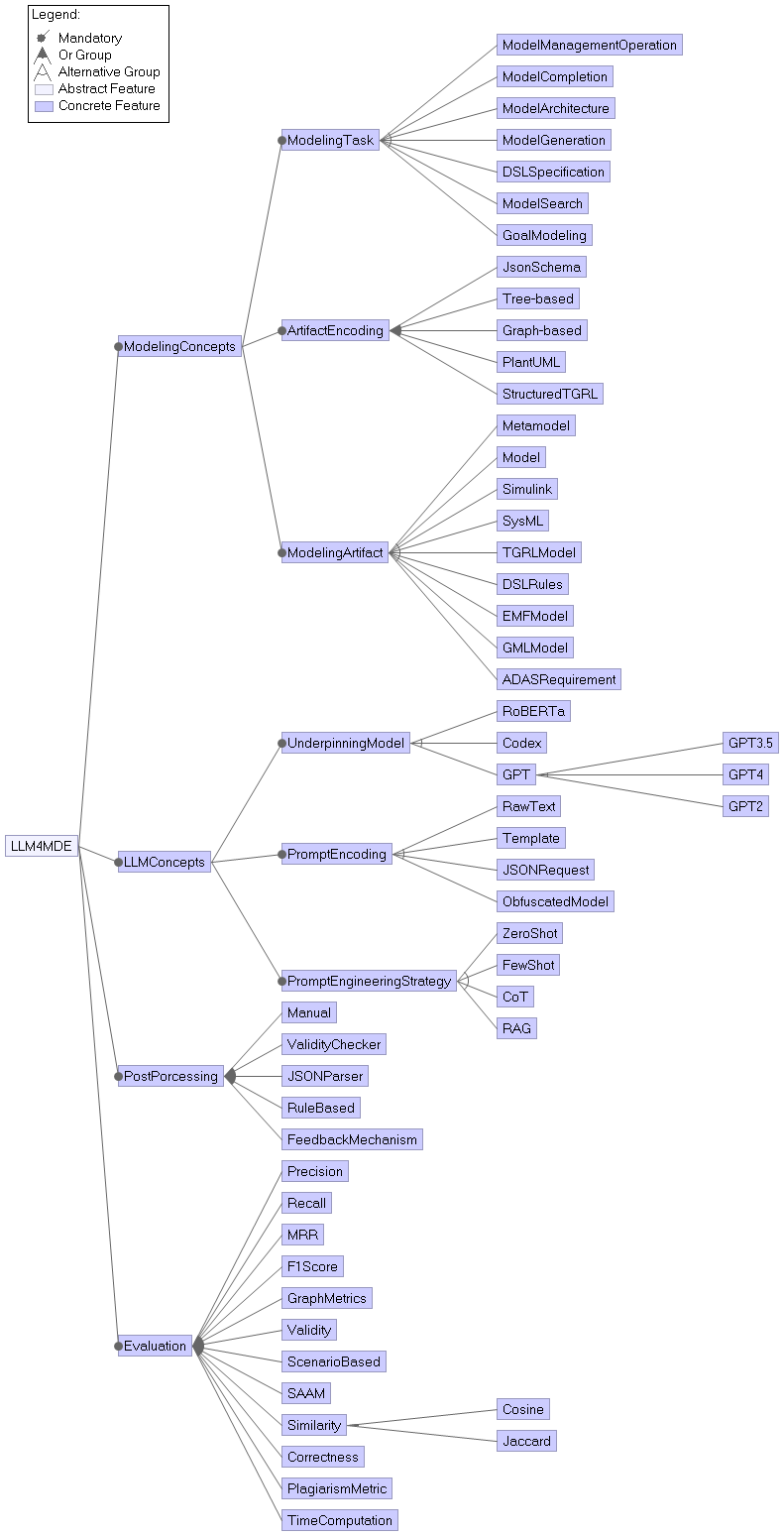}
	\caption{Feature models representing state-of-the-art approaches employing \Ls and MDE.}
	\label{fig:features}
\end{figure*}

From the analyzed approaches, we constructed a feature model depicted in Figure \ref{fig:features}. This model encapsulates the fundamental concepts of both fields, \ie \Ls and MDE. The devised feature model provides a technical visualization of the essential elements and relationships between \Ls and MDE, offering insights into their interconnected functionalities and constituting elements. Concerning the \textit{MDEConcepts}, the identified \textit{Input Artifact} needs to be encoded by adopting the proper \textit{artifact Encoding}, \ie standard format that can be processed by LLMs. It is worth mentioning that there are only three types of model artifacts supported by the existing literature, \ie domain models, metamodels, and Simulink models. Concerning \textit{LLMConcepts}, we map the generic concepts discussed in Section \ref{sec:background} to the actual approaches identified in Table \ref{tab:comparison}. %In particular, 
\revised{We report that \textit{PromptEncoding} plays an important role since it is necessary to guide the generation of the wanted \textit{Input Artifact}. Different approaches show that using a prompt template compared to \textit{RawText} can improve the quality of the generated modeling artifacts. Notably, only one of the examined approaches employs RAG as a strategy, \ie Apvrille and Sultan \cite{modelsward24}. This means that the majority of identified works focus more on  zero and few-shots prompting. Noteworthy, few approaches handle the post-processing phase by relying on two main techniques, \ie tailored parsers or iterative feedback. Concerning the evaluation, we report that accuracy metrics are used mostly to evaluate prediction tasks, \eg model completion or model management operations, while tasks that require human reasoning, \eg goal modeling or model architecture, have been evaluated using scenarios or use cases. }

\revised{Concerning MDE4LLM, Cariso and Cabot \cite{Models23PE} proposed a domain-specific language (DSL), called Impromptu, to generate prompts in a platform-independent way. The system allows users to define the prompt sketch, customize it for the application domain, and validate it using a dedicated code generator. To evaluate the Impromptu capabilities, the authors generate platform-specific fine-tuned prompts for two platforms, \ie, Midjourney and Stable Diffusion. The results demonstrate that the proposed DSL succeeded in supporting the image.to-text systems, although in-depth evaluation is required to support more complex LLMs. Since it is the only work that supports MDE4LLM, we did not insert it in the table to avoid an unfair comparison, \ie the identified feature refers to LLM approaches to support modeling tasks.}

%\DDR{Evaluation metrics are presented in Table 2 and in the feature diagram, but are completely missing in the descriptive text.}

%\begin{shadedbox}
%\textbf{Remark:} Overall, most of the considered applications of \Ls to MDE tasks make use of basic prompt techniques and GPT models. Notably, only one of the existing approaches considers retrieval augmented generation (RAG) to enhance the accuracy of the underpinning models.  %Considering the covered tasks, we report that LLMs have been mostly used to support domain modeling and model completion. 
%\end{shadedbox}

\begin{shadedbox}
	\revised{\textbf{Answer to RQ$_1$:} LLMs are increasingly being used in MDE tasks such as model completion, and generation. While models like GPT-3 and GPT-4 have been successfully applied, the field is still emerging, with significant opportunities in developing more sophisticated integration methods such as Retrieval-Augmented Generation (RAG) and knowledge graph-based approaches.}
\end{shadedbox}

\begin{shadedbox}
	\revised{\textbf{Answer to RQ$_2$:} Prompt engineering, few-shot and zero-shot learning methods  are commonly used to adapt LLMs to specific MDE tasks without extensive fine-tuning. Some approaches have explored the use of LLMs in conjunction with MDE tools, through iterative feedback loops and post-processing.}
\end{shadedbox}

%\begin{shadedbox}
%	\textbf{Remark \#2:} Notable applications in LLMs are limited to domain modeling, thus left uncovered a wide range of MDE tasks, spanning from model transformation to DSL specification. 
%\end{shadedbox}
%

\section{Research agenda}
\label{sec:agenda}
\begin{figure*}[h!]
	\centering
	\includegraphics[width=0.8\textwidth]{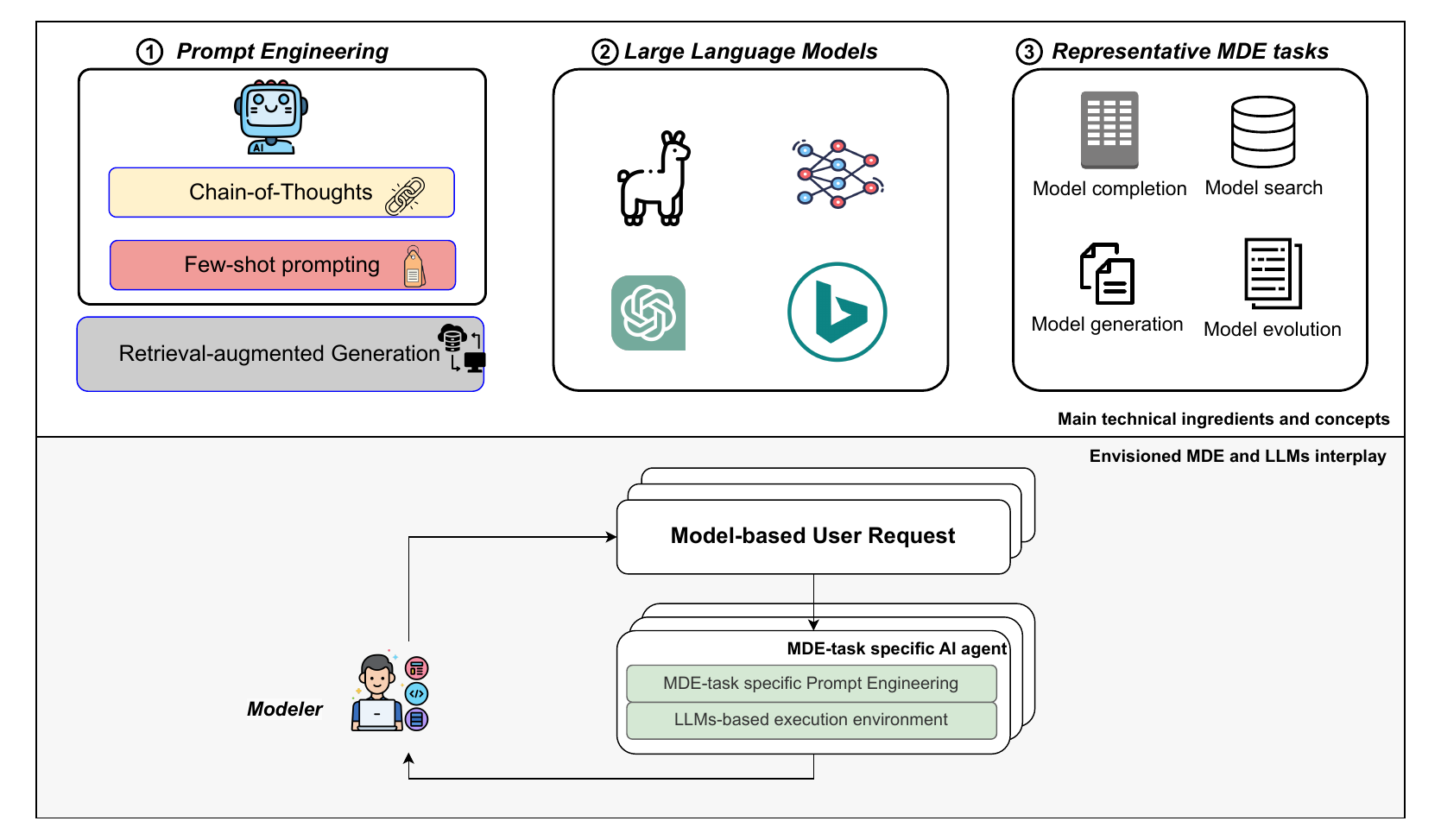}
	\caption{\revised{Envisioned interplay of MDE and \Ls.}}
	\label{fig:Architecture}
	\vspace{-.2cm}
\end{figure*}

The integration of \Ls  in the realm of MDE introduces a transformative dimension to the established concepts of \textit{abstraction} and \textit{automation}. Traditionally, MDE has been centered around abstracting target platforms and providing automation to simplify the engineering of complex systems. This has proven effective in managing the intricacies of diverse platforms and streamlining development processes. However, with the advent of \Ls, MDE can now extend its capabilities to support the adoption of single \Ls and the interactions of several \Ls. 

\revised{\Ls differ from other AI-based techniques in several key aspects including \textit{generative capabilities} and \textit{pre-training on vast corpora} by disclosing several opportunities for supporting automation in MDE tasks. In particular, the generative ability of \Ls distinguish them from other AI techniques that might focus on classification, regression, or other predictive tasks. Moreover, the extensive pre-training of \Ls allow them to perform well on various tasks with minimal fine-tuning. In contrast, other AI techniques often require task-specific training data and significant feature engineering.}

Akin to the concepts of Software Engineering for AI (SE4AI) and AI for Software Engineering (AI4SE), it is imperative for the MDE community to actively engage with two distinctive but interrelated directions: MDE4LLM and LLM4MDE. They represent a bidirectional collaboration aimed at advancing the integration and utilization of \Ls in diverse domains while leveraging the capabilities of \Ls to enhance various MDE tasks.
In this section, we dare draft a research agenda organized with respect to the two identified directions.

\subsection{\textit{MDE4LLM}: Supporting \Ls adoption with MDE}

In the MDE4LLM direction, the focus is on supporting the adoption of \Ls in various domains by means of model driven techniques and tools. Envisioning a multitude of task-specific \Ls, the community should actively contribute to the training and utilization of these models to support a broad spectrum of tasks, extending beyond traditional software engineering domains. Recognizing that the applications of \Ls span diverse fields, the MDE community should reinforce the multidisciplinary attitude. While the immediate applications may not be limited to software engineering tasks, the community's expertise in modeling and abstraction can significantly contribute to the development and effective use of \Ls in different research and application domains.

\smallskip
\noindent
\textit{Abstraction in the context of \Ls:} MDE has historically focused on abstracting the intricacies of target platforms through domain specific modeling languages. Convergence of domain-specific languages, \Ls, and prompt engineering emerges as a promising trajectory, particularly in training and using task-specific language models. Prompt engineering becomes an integral part of this research, facilitated by repositories of context-aware and domain-specific languages.

\smallskip
\noindent
\textit{Automation with multitudes of task-specific \Ls:}  In the traditional MDE sense, a \textit{platform} refers to a specific technology stack or execution environment. With the integration of \Ls, the notion of a platform expands to include the collaborative ecosystem of multiple language models working together in a multi-agent system. This conceptual shift broadens the scope of MDE's target platforms. 

\smallskip
\noindent
\textit{Envisioned interplay of MDE and \Ls:} As illustrated in Fig. \ref{fig:Architecture}, the symbiotic integration of \Ls and MDE  technologies and tools holds the potential for reciprocal benefits. Building upon our earlier discussion, we anticipate a paradigm shift towards the proliferation of task-specific AI agents fueled by \Ls. This departure from large, generalized agents, which entail resource-intensive development and training processes, offers notable advantages in terms of both time and cost-effective\-ness.

In this envisioned landscape, \Ls transition from playing the role of generic execution environments to empowering MDE-task specific AI agents. These specialized agents are tailored to perform distinct model management operations, including but not limited to model evolution, model comparison, domain modeling, generation of training and testing data, and model completion. Users can increase their efficiency and relevance in development processes by aligning AI agents with specific MDE tasks. Crucially, the input to these task-specific agents is provided through specifications adhering to domain-specific languages (similarly to what was proposed by \textit{Claris\'o} \etal \cite{Models23PE}), further enhancing the precision and applicability of the collaborative interplay between \Ls and MDE technologies. In other words, low-code environments \cite{DBLP:journals/sosym/RuscioKLPTW22} will democratize the usage of AI agents that are specific for the application domains of interests.

\subsection{\revised{\textit{LLM4MDE}: \Ls-based automation for MDE tasks}}

In the LLM4MDE direction, the community should continue the work initiated with automated modeling assistants, domain analysis based on neural networks, and deep learning technologies. 
%\Ls are positioned as automation components to enhance and support various MDE tasks, emphasizing their role as intelligent collaborators in the model-driven ecosystem. 
The scope of LLM4MDE extends beyond conventional software engineering tasks. \Ls can be leveraged as tools to automate and assist in a wide range of MDE activities, from requirements elicitation to model transformation, providing valuable insights and augmenting the capabilities of modelers.

\revised{A cross-cutting concern that involves all the aspects discussed later in this section is the mitigation of hallucinations in LLMs. Hallucinations occur when LLMs produce responses that are not relevant to the input context, resulting in nonsensical, incorrect, or useless outputs. It is then crucial to address this issue to ensure the reliability and effectiveness of LLMs in MDE tasks. Various strategies, such as fine-tuning on domain-specific data, integrating additional context or constraints into the generation process, and implementing post-generation filtering techniques, can help mitigate hallucinations. Ensuring that LLMs provide accurate and contextually appropriate outputs is essential for their successful integration into MDE workflows. Moreover, designing and deploying LLMs applications with the user-centered principle will help improve the velocity and flexibility of the development process. Once the goals and needs of the system’s end-users have been placed at the center of software development, they will allow developers to deliver software with appropriate usability~\cite{BRHEL2015163}.}

\smallskip
\noindent
\textit{Enhanced modeling assistance:} \revised{Being built on top of %on the foundation of 
automated modeling assistants and domain analysis based on neural networks, research in this direction should focus on advanced recommendation algorithms, combining LLM capabilities with existing MDE knowledge to provide accurate and context-aware suggestions for modelers. In particular, besides the research already done so far (see Section \ref{sec:SLR}), further investigations are needed to enable personalized recommendations based on individual modeling styles and preferences. For instance, RAG techniques can be adopted to enable the generation of recommendations that not only consider the inherent characteristics of modeling artifacts but also adapt to the specific context and preferences of the modeler.}

%\color{blue}
\smallskip
\noindent
\textit{Dealing with hallucinations:} We do not expect to provide an exhaustive list of methods to mitigate hallucinations, rather than, we anticipate that there are at least the following ones: %possible countermeasures:

\begin{itemize}
	
	\item \textit{Training and fine-tuning with high quality datasets}: Using high-quality, well-curated training datasets reduces the chances of the model learning incorrect information. Including diverse perspectives and accurate information helps the model to provide more accurate outputs.
	
	\item \textit{Prompt Engineering}: Designing prompts that are clear and specific can guide the model toward generating more accurate responses. Moreover, providing context within prompts will help the model understand the scope and constraints, aiming to reduce the likelihood of generating off-topic or fabricated information.
	
	\item \textit{Users' feedback}: Incorporating feedback from users is a means to continuously improve the model’s performance and reduce hallucinations over time. Moreover, allowing users to query the model iteratively and correct or clarify information can help mitigate hallucinations.
	
\end{itemize}

\color{black}

\smallskip
\noindent
\textit{Automated model generation:} This line of research is to enhance the automation capabilities within MDE by extending to the automatic generation of models or model elements based on insights from LLMs. As discussed in the previous sections, different techniques have been already proposed to synthesize models \cite{varro2018towards} or mutate existing ones according to user specified characteristics \cite{gomez-abajo_wodel_2016} \eg to evaluate new model management tools on varying model data sets. Additionally, \Ls can assist in generating or mutating models by analyzing textual requirements and considering the specificity of the application domain of interest. Thus, for example when asked to create mutants of a given input model, instead of generating model elements named with random strings, the application domain (such as medical or industrial) will be considered to generate appropriate elements having names that make sense for the domain of interest. \revised{In this respect, prompt engineering can be used to craft effective queries or input patterns, so as %It is related to forming the input text, so as to yield the desired response or behavior from the model. 
to provide the model with context and constraints that steer its output towards the desired results.}

\smallskip
\noindent
\textit{Ethical and responsible use of \Ls in MDE:} As LLMs become integral to MDE workflows, the ethical implications of their usage come into focus \revised{as stated by recent research \cite{basta-etal-2019-evaluating,gehman-etal-2020-realtoxicityprompts,10298519,NEMANI2024100047}}. Research efforts should aim to examine issues related to bias, fairness, and transparency, proposing guidelines and best practices for the responsible integration of \Ls. The focus has to extend beyond traditional software engineering tasks, ensuring ethical considerations in diverse application domains. The relevance of such topics has been made popular, \eg by infamous incidents in the recruitment instrument employed by Amazon\cite{guardian2018amazonai} and the criminal recidivism predictions made by the commercial risk assessment software COMPAS \cite{propublica2016compas}.

\smallskip
\noindent
\textit{Benchmarking and evaluation metrics:}  \revised{To facilitate the integration of LLMs into MDE workflows, researchers and practitioners can already exploit several tools and frameworks. For instance, Hugging Face Transformers\footnote{https://huggingface.co/docs/transformers/index} is a widely used library that offers access to a vast array of pre-trained LLMs, such as GPT, BERT, and more, which can be fine-tuned or used directly in MDE applications. OpenAI API\footnote{https://openai.com/} allows access to LLMs like GPT-3 and GPT-4, enabling developers to implement these models in MDE scenarios. Over the last few years, some datasets (\eg MAR \cite{lopez2020mar}, and ModelSet\footnote{https://models-lab.github.io/blog/2021/modelset/}) have been defined by the MDE community, and they can be used to further train or evaluate LLMs within the context of MDE.} 

To gauge the effectiveness of \Ls integration in MDE tasks, establishing standardized benchmarks becomes crucial. Metrics including model accuracy, efficiency, and adaptability go beyond traditional evaluation criteria. \revised{Moreover, qualitative metrics such as fairness, robustness should also be taken into account. In this respect, human evaluation is an essential step to validate the performance of an LLM. Such an evaluation, apart from a conventional quality measurement, can also help reveal the Helpfulness, Honesty, and Harmlessness of an LLM~\cite{Askell2021AGL}. In fact,  a manual evaluation reflects better the actual application scenario, and thus it has the potential to yield more comprehensive and accurate feedback~\cite{10.1145/3641289}.} This research direction should aim to provide a comprehensive framework for evaluating \Ls performance within the requirements of MDE scenarios. Similarly to what has been done to support a disciplined comparison of ML methods for a particular MDE task \cite{LopezRCR22}, there will be the need for frameworks to support the quality assessment of contents generated by \Ls and even to compare those that are produced by different models from the same queries.

\smallskip
\noindent
\textit{Scalability and resource efficiency:} Scalability is a crucial consideration in deploying \Ls for large-scale MDE projects. Research needs to investigate methods to enhance scalability while optimizing resource utilization. Thus, techniques for deploying \Ls in resource-cons\-trained environments must be explored, ensuring accessibility across a broad spectrum of MDE applications. Here, we need support to suggest the most energy/re\-source-efficient technique that can be exploited for the problem at hand. An early analysis is preferable instead of always using the most resource-consuming technologies. Of course, this can be done when the user can accept the reduced accuracy price. In other words, it is necessary to devise methodologies for early analysis and assessment of available models and techniques to identify the most suitable options for the given MDE task. Such proactive approaches will allow for informed decision-making, balancing resource constraints with the desired level of accuracy.

\smallskip
\noindent
\textit{Security and robustness:} Security is a crucial consideration when integrating \Ls within MDE workflows. It is essential to assess potential vulnerabilities and propose mitigation strategies to ensure the integrity of the model-driven ecosystem. The research direction should focus on exploring techniques to enhance the robustness of \Ls against adversarial attacks within MDE tasks. For instance, recently GitHub faced an attack that resulted in the creation of millions of code repositories containing obfuscated malware \cite{arstechnica2024github} These malicious repositories are clones of legitimate ones, making them challenging to distinguish. An unknown party automated a process that forks legitimate repositories, resulting in millions of forks with names identical to the originals but containing payloads wrapped under seven layers of obfuscation. Therefore, it is necessary to develop techniques and tools for detecting malicious sources when training language models for MDE tasks or any other software engineering purposes.

\smallskip
\noindent
\textit{Long-term impact assessment:} The long-term impact of adopting \Ls in MDE practices is a crucial aspect of research. Thus, it is necessary to perform studies to assess, \eg modeler productivity, and overall product quality. Key success indicators and performance metrics have to be identified to measure sustained benefits, providing insights into the impact of \Ls integration in MDE over time. 

\smallskip
\noindent
\textit{Educational assistance:} Research can be done to employing \Ls for teaching. In particular, the goal is integrating \Ls into educational environments to teach and assist new modelers in their tasks. This concept aligns with recommender systems but focuses on less granular concepts, providing educational support at a broader level. \Ls can offer explanations, context, and guidance tailored to the educational needs of novice modelers.  To this end, the community might consider training \Ls using reviewed sources such as SWEBOK \cite{computer2024software}, SLEBOK,\footnote{\url{https://slebok.github.io/}} and thereby contribute to the development of a comprehensive Model-Based Software Engineering Body of Knowledge \cite{MDEBOK2019}.

\begin{figure}
	\centering
	\includegraphics[width=\linewidth]{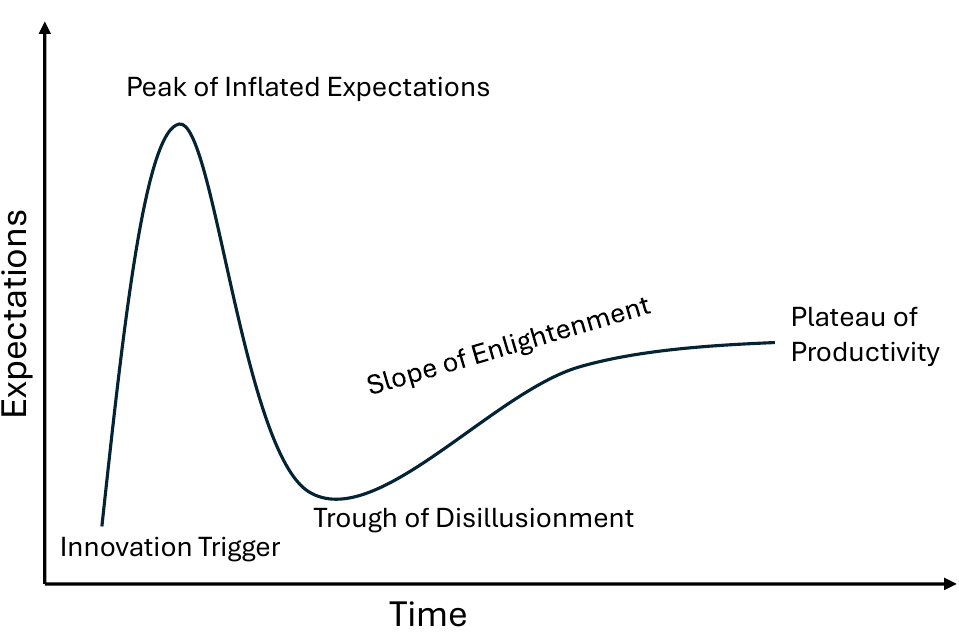}
	\caption{The Gartner Hype Cycle.}
	\label{fig:tech-expectations}
\end{figure}

\smallskip
\noindent
\textit{Final thoughts:} The hype around using \Ls is showing their big potential in many areas. But like any new technology, the initial excitement might fade as we learn more about what they can and can not do. In particular, mirroring the Gartner Hype Cycle\footnote{\url{https://en.wikipedia.org/wiki/Gartner_hype_cycle}} (see Fig. \ref{fig:tech-expectations}), \Ls may undergo phases of inflated expectations, followed by a clearer understanding of their real-world applications and constraints.

We believe that \Ls represent a relevant technology to improve the \emph{automation} aspect of model-driven engineering. However, while \Ls-based automation can benefit MDE processes, the human element remains indispensable. In particular, 
% for mitigating errors, ensuring accountability, and meeting certification requirements. 
%
as discussed in Section \ref{sec:background}, \Ls may produce outputs that deviate from the desired context or contain inaccuracies. Human intervention is crucial to identify and rectify such instances, ensuring the quality and reliability of model-driven artifacts. Moreover, in complex engineering tasks, particularly those involving critical systems, human oversight remains indispensable for accountability. Human-in-the-loop systems enable traceability and accountability, ensuring that decisions made by \Ls align with regulatory standards. Finally, industries such as healthcare, aerospace, and automotive adhere to stringent certification standards. Human involvement in the loop is still necessary for the validation and certification processes, ensuring compliance with regulatory requirements and safety standards.

\section{Related work}
\label{sec:relatedwork}

%  and David Lo LLM survey in SE.  
In recent years, the research landscape surrounding the integration of LLMs with SE has witnessed remarkable activity. This section reviews some of the most notable studies in this topic. 

%In \cite{Martinez2022} the authors conducted 
A systematic mapping study~\cite{Martinez2022} to analyze 248 studies from January 2010 to March 2020 reveals that the most explored SE properties of AI-based systems are dependability and safety. The study identifies various SE approaches for AI-based systems, categorized according to the SWEBOK areas, with a focus on software testing and quality, while maintenance aspects appear neglected. Data-related challenges are recurring, providing valuable insights for researchers, practitioners, and educators to understand the current state-of-the-art, address research gaps, and bridge the knowledge divide between SE and AI in curricula.

Hou \etal \cite{Hou_Zhao_Liu_Yang_Wang_Li_Luo_Lo_Grundy_Wang_2024} focused on the application of \Ls in Software Engineering (from 2017 to 2023). Firstly, the study categorizes different \Ls used in SE tasks, describing their features and applications. Secondly, it analyzes data collection, preprocessing methods, and the importance of well-curated datasets for successful implementation. Thirdly, it investigates strategies for optimizing and evaluating LLM performance in SE. Lastly, it examines specific SE tasks where LLMs have demonstrated success, highlighting their practical contributions. The review aims to provide a comprehensive understanding of the current state-of-the-art, identify research gaps, and suggest promising areas for future study in the intersection of LLMs and SE.

Similarly, Fan \etal \cite{Fan_Gokkaya_Harman_Lyubarskiy_Sengupta_Yoo_Zhang_2023} presented a survey on \Ls in software engineering and highlights the research challenges in applying \Ls to address technical issues faced by software engineers. LLMs are known for their innovative and generative capabilities, which have a significant impact on various SE activities such as coding, design, requirements, repair, refactoring, performance improvement, documentation, and analytics. However, the emergence of these properties also poses significant challenges in accurately identifying solutions and addressing issues like hallucinations. The survey emphasizes the importance of hybrid techniques that combine traditional SE approaches with \Ls to ensure the development and deployment of reliable, efficient, and effective LLM-based SE solutions.

In the MDE community, Combemale \etal \cite{combemale_chatgpt_2023} discussed how a large language model like ChatGPT can be used in software development, particularly in creating models that represent software systems. The authors explored different scenarios, from fully automated generation of code from requirements to using ChatGPT as an assistant for human modelers. They acknowledged the challenges of ensuring reliable and trustworthy results from AI-generated models and the need for large libraries of existing models for ChatGPT to learn from.

Di Ruscio \etal \cite{DiRuscio2023} elaborated on %discussed 
the use of model-driven engineering and machine learning  techniques to support the management of modeling ecosystems. 
%MDE promotes the use of models to support the development of complex systems, while ML algorithms can extract meaningful patterns from data. The MDE community has investigated the use of ML/DL to facilitate the automated classification of model repositories and develop recommender systems. 
The paper identifies and discusses possible lines of research to explore the adoption of existing machine learning techniques to enhance the management of modeling ecosystems.

In a recent paper \cite{camara_assessment_2023}, the authors elaborated on the potential of \Ls, such as Copilot and ChatGPT, to revolutionize software development. The paper examines the current capabilities of ChatGPT for modeling tasks and assisting modelers, and identifies several shortcomings, including syntactic and semantic deficiencies, lack of consistency in responses, and scalability issues. The paper provides suggestions on how the modeling community can help improve the current capabilities of ChatGPT and future \Ls for software modeling.

%The work done by Chen \etal \cite{PromptingFinetuningComparative2021} discusses the use of taxonomies, which are hierarchical relationships between entities used in software modeling and natural language processing. Though constructing taxonomies manually is time-consuming and costly, recent studies have shown that large language models can be guided effectively by user inputs. The paper presents a general framework for taxonomy construction, taking into account structural constraints. The comparison of the prompting and fine-tuning approaches shows that prompting performs better than fine-tuning, even without explicit training on the dataset. However, taxonomies generated by the fine-tuning approach can be easily post-processed to meet all constraints, whereas handling violations of taxonomies produced by the prompting approach can be challenging.
%\cite{10173990} \cite{Models23AM}, \cite{Models23PE}. 

\section{Conclusion and future work}
\label{sec:conclusion}
%\PN{Conclusions go here.}

%In this paper, we provide an overview of current applications of Language Large Models (LLMs) in MDE, emphasizing their role in automating tasks like model repository classification and developing advanced recommender systems. The paper also outlines the technical considerations for seamlessly integrating LLMs into the MDE workflow, offering a practical guide for researchers and practitioners.

In contrast to broader research on Large Language Models   in Software Engineering, our paper focuses on the specific synergy between LLMs and Model-Driven Engineering. We explored how LLMs automate tasks unique to MDE, like model repository classification and advanced model recommenders. The paper also outlines the technical considerations for seamlessly integrating LLMs into MDE workflows, offering a practical guide for researchers and practitioners. This paper proposed also a targeted research agenda, identifying challenges and opportunities for leveraging LLMs in MDE and vice-versa. This  roadmap contributes to evolving MDE practices and offers a forward-looking perspective on the transformative role of Large Language Models in software engineering and model-driven practices.

\vspace{.2cm}

{\noindent\bf Acknowledgments}. This work has been partially supported by the EMELIOT national research project, which has been funded by the MUR under the PRIN 2020 program (Contract 2020W3A5FY). The work has also been partially supported by the European Union - NextGenerationEU through the Italian Ministry of University and Research, Projects PRIN 2022 PNRR ``FRINGE: context-aware FaiRness engineerING in complex software systEms'' grant n. P2022553SL. We acknowledge the Italian ``PRIN 2022'' project TRex-SE: ``Trustworthy Recommenders for Software Engineers,'' grant n. 2022LKJWHC. \revised{We thank the anonymous reviewers for their useful comments and suggestions that helped us improve our manuscript.} 

%\textbf{ADD ALSO FRINGE}. % AND TREXSE

%Work partially supported by the EMELIOT national research project, which has been funded by the MUR under the PRIN 2020 program (Contract 2020W3A5FY). 

\bibliographystyle{spmpsci}      % mathematics and physical sciences
\bibliography{main}   % name your BibTeX data base

\end{document}